\documentclass[fleqn,usenatbib]{mnras}

\usepackage{newtxtext,newtxmath}

\usepackage[T1]{fontenc}

\DeclareRobustCommand{\VAN}[3]{#2}
\let\VANthebibliography\thebibliography
\def\thebibliography{\DeclareRobustCommand{\VAN}[3]{##3}\VANthebibliography}

\usepackage{graphicx}	
\usepackage{amsmath}	
\usepackage{array}
\usepackage{multirow}
\usepackage{threeparttable,booktabs}

\title[Structural stability of asteroid 2016 HO$_{3}$]{Structural stability of China's asteroid mission target 2016 HO$_{3}$ and its possible structure}

\author[B. Cheng et al.]{
Bin Cheng$^{1}$\thanks{E-mail: bincheng@mail.tsinghua.edu.cn}
and Hexi Baoyin$^{1,2}$
\\
$^{1}$Tsinghua University, Beijing 100086, China\\
$^{2}$Inner Mongolia University of Technology, Inner Mongolia, 010051, China
}

\date{Accepted XXX. Received YYY; in original form ZZZ}

\pubyear{2023}

\begin{document}
\label{firstpage}
\pagerange{\pageref{firstpage}--\pageref{lastpage}}
\maketitle

\begin{abstract}
Asteroid 2016 HO$_3$, a small asteroid (<60 m) in super fast rotation state ($\sim$28 min), and is the target of China's Tianwen-2 asteroid sample-return mission.
In this work, we investigate its structural stability using an advanced soft-sphere-discrete-element-model code, DEMBody, which is integrated with bonded-aggregate models to simulate highly irregular boulders.
The asteroid body is numerically constructed by tens of thousands particles, and then is slowly spin up until structural failure.
Rubble piles with different morphologies, grain size distributions and structures are investigated.
We find a 2016 HO$_3$ shaped granular asteroid would undergo tensile failure at higher cohesive strengths as opposed to shear failure in lower strengths, regardless of its shape and constituent grain size ratio.
Such a failure mode transition is consistent with the priority between the Maximum Tensile Stress criterion criterion and the Drucker–Prager criterion.
Therefore, previous works that solely considered the Drucker–Prager failure criterion have underestimated the minimal cohesion strength required for fast-rotating asteroids.
We predict that the high spin rate of asteroid 2016 HO$_3$ requires a surface cohesion over $\sim$1 Pa and a bulk cohesion over $\sim$10--20 Pa.
Through comparing these strength conditions with the latest data from asteroid missions, we suggest a higher likelihood of a monolithic structure over a typical rubble pile structure. However, the possibility of the latter cannot be completely ruled out.
In addition, the asteroid's surface could retain a loose regolith layer globally or only near its poles, which could be the target for sampling of Tianwen-2 mission.
\end{abstract}




\section{Introduction}
Asteroids, remnants from the planet formation era, have garnered intense interest in deep-space exploration due to their potential to shed light on the primordial processes that shaped our Solar System, as well as their prospective value in space resource utilization. 
Among these small celestial bodies, Earth's quasi-satellites, in their unique orbital resonance with Earth, represent a unique subset \citep{connors2004discovery}. 
Asteroid 2016 HO$_3$, or, Kamo`oalewa, is a unique one in this category \citep{de2016asteroid}.
It is the most stable of Earth's five known quasi-satellites, and has been a close companion to our planet over millions of years \citep{castro2023lunar}.
Orbiting the Sun while remaining close to the Earth, 2016 HO$_3$ provides a rare opportunity to study a near-Earth object (NEO) with relative ease of access.
\citet{sharkey2021lunar} found 2016 HO$_3$ has a tiny size about 36--60~m in diameter, which is smaller than all other asteroids explored by space missions.
Moreover, it rotates at an unusually rapid speed (28.3 min), much faster than most asteroids.
Its reflectance spectrum diverges from typical NEO spectra \citep{demeo2009extension}, including S-, Q-, and C-types, but instead bears a striking resemblance to Lunar-like silicate material \citep{sharkey2021lunar}.
This raises the fascinating hypothesis that this quasi-satellite could originate as impact ejecta from the lunar surface, a theory that further enhances its scientific appeal.
These unique characteristics make 2016 HO$_3$ an ideal target for in-situ investigation to advance our understanding of asteroid dynamics, composition, and potential resource utilization.
In this context, China's Tianwen-2 mission is set to rendezvous with 2016 HO$_3$ and land on the asteroid to collect a sample of surface regolith \citep{zhang2019zhenghe}.
The spacecraft will, for the first time, employ the anchor-and-attach method to attempt collection of a sample from the asteroid \citep{wang2023study}, which is different from the touch-and-go method used by Hayabusa2 and OSIRIS-REx missions \citep{bierhaus2018osiris,sawada2017hayabusa2}.
Therefore, accurately assessing the strength and structure of 2016 HO$_3$ is crucial for the mission's success, influencing key mission decisions from the design of landing strategies to the methods of sample collection.

\begin{figure*}
	\includegraphics[clip,trim=0mm 0mm 0mm 0mm,width=\textwidth]{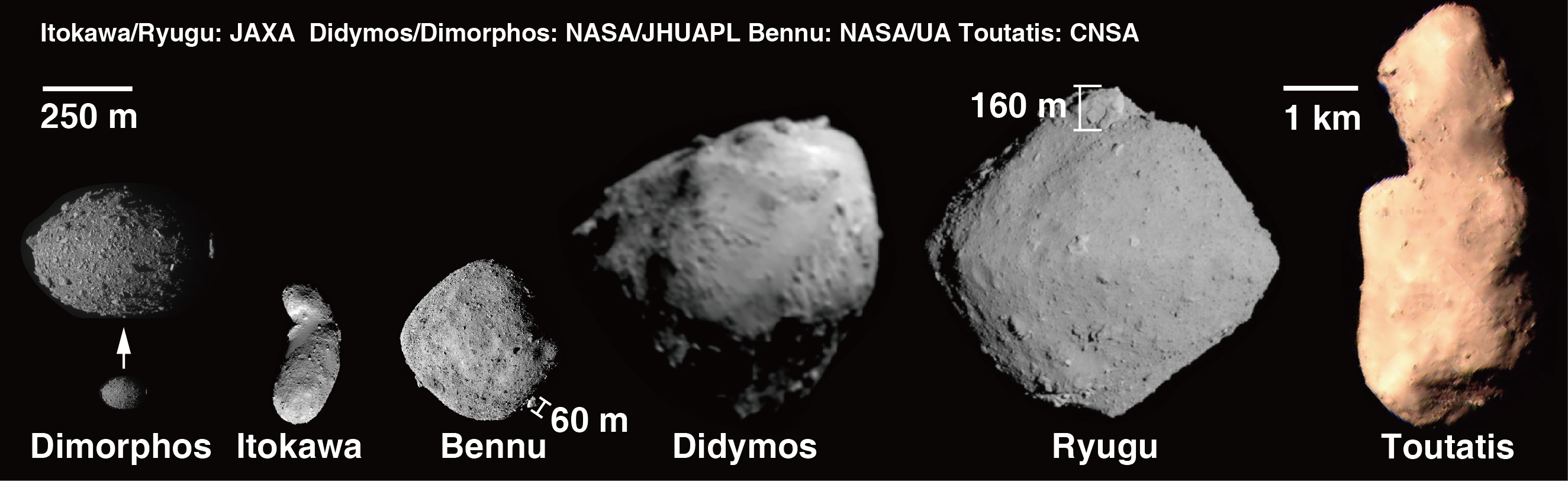}
    \caption{Asteroids that have been explored by space missions. These asteroids >200 m and <10 km in diameter all possess rubble pile structures. But asteroid 2016 HO$_3$, being only tens of meters in diameter, falls outside this range of direct observation.}
    \label{fig:asteroids}
\end{figure*}

Our direct understanding of asteroid strength and structure has predominantly been shaped by in situ spacecraft observations. 
These different space missions have consistently confirmed that asteroids >200 m and <10 km in diameter possess rubble pile structures, i.e., an unorganized collection of rubbles held together by their self-gravity. This pattern spans a range of sizes as shown in Fig.~\ref{fig:asteroids}, from the approximately 2-km-sized asteroid Toutatis \citep{huang2013ginger}, to Ryugu at around 1000 m \citep{sugita2019geomorphology}, Didymos at about 800 m \citep{cheng2023momentum}, Bennu near 500 m \citep{lauretta2019unexpected}, Itokawa at roughly 300 m \citep{fujiwara2006rubble}, down to the 150-m-sized Dimorphos \citep{cheng2023momentum}. This observation across various sizes seems to reinforce the rubble pile hypothesis as a common structural characteristic among smaller asteroids.
Numerical simulations of collisional formation of asteroids lend support to this perspective \citep{michel2013collision}; they suggest that in most cases where an asteroid is disrupted by a collision with another object, the velocities at which fragments are ejected are generally slow enough to allow most of these fragments to reaggregate under mutual gravity, consequently leading to the formation of progeny asteroids with rubble pile structures.
However, these observed asteroids have all been over a hundred meters in size, which means asteroid 2016 HO$_3$, being only tens of meters in diameter, falls outside this range of direct observation. 
Direct exploration of such smaller-scale asteroids has not yet been conducted. 
However, on larger rubble pile asteroids, we have indeed observed numerous monolithic boulders in the tens of meters range.
For instance, the asteroid Bennu hosts the 60-m Benben Saxum with sinuous cracks, which is thought to be the remnants of the parent asteroid post-impact \citep{walsh2019craters,dellagiustina2021exogenic}.
Previous asteroid disruption simulations by \cite{benz1999catastrophic} have found that smaller asteroids fall within a strength-dominated regime, where gravitational forces are insufficient to reaggregate fragments post-collision, facilitating the formation of monolithic asteroids.
However, the precise threshold size for this transition remains unclear.
Therefore, whether 2016 HO$_3$ is more akin to a scaled-down version of the larger rubble pile asteroids or more similar to the monolithic boulders like Benben Saxum remains an open question. 
This uncertainty poses a significant risk to the Tianwen-2 mission's anchor-and-attach landing operation.

\begin{figure*}
	\includegraphics[clip,trim=0mm 0mm 0mm 0mm,width=0.8\linewidth]{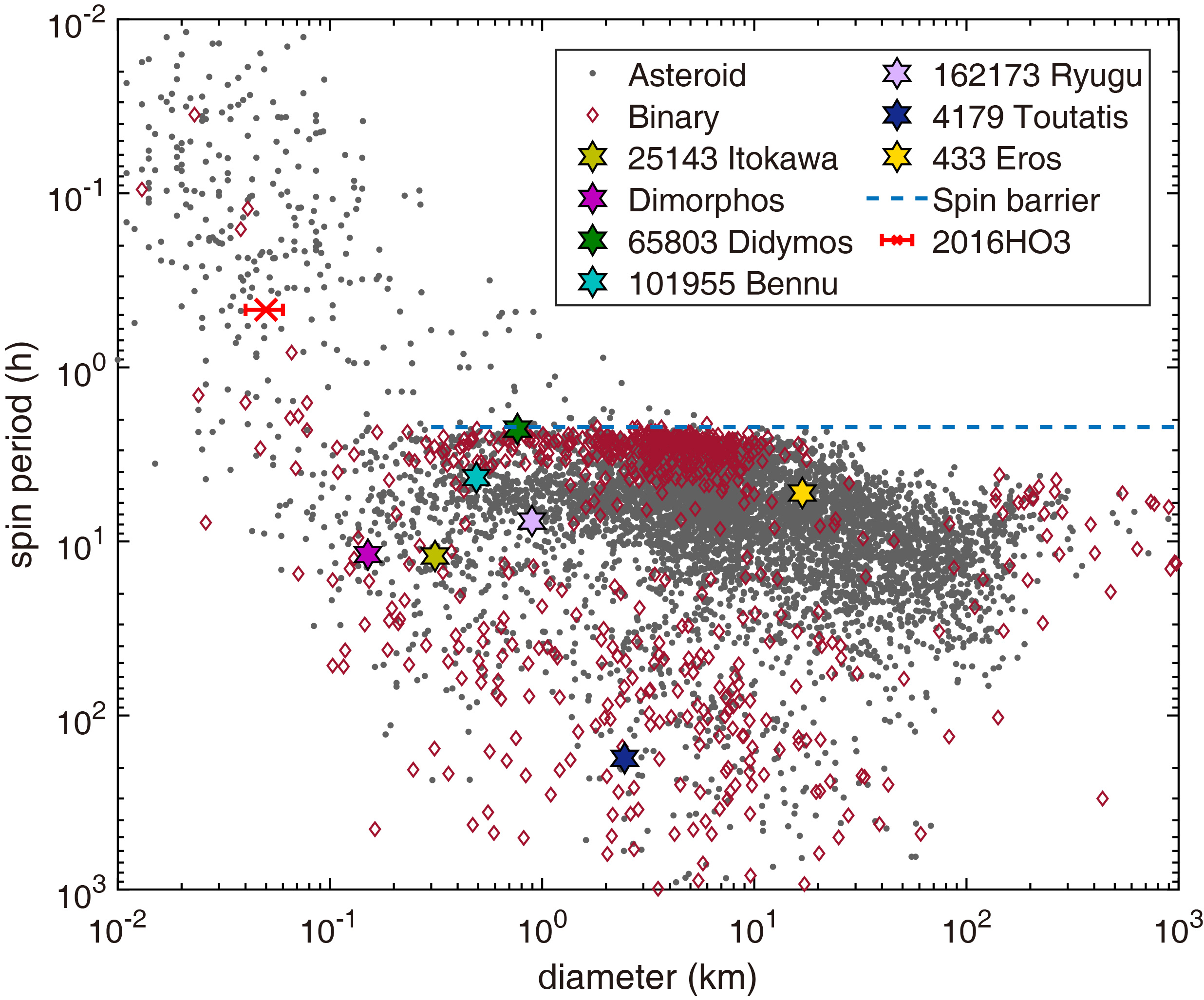}
    \caption{Asteroid size vs. spin period, with binaries and objects explored by space missions called out specifically. Asteroid 2016 HO$_3$ spins much faster than the spin barrier of ~2.2 h, but its size is also smaller than the range for which the spin barrier is typically relevant (>200 m).}
    \label{fig:spinsize}
\end{figure*}

The mass, spin and shape as a combined constraint offers another powerful tool to estimate the structure and strength of asteroids indirectly \citep{walsh2018rubble}.
Figure \ref{fig:spinsize} shows asteroid diameters and spin periods taken from the updated version of the Asteroid Lightcurve Database \citep{warner2009asteroid}.
A noticeable trend in this data is the absence of very rapidly rotating bodies, indicative of a spin barrier at ~2.2 h. This rate is similar to the rotation speed at which free particles could detach from the surface of a spinning body. 
This finding generally indicates some monolithic or strength-dominated bodies with spin rates much faster than the spin barrier of ~2.2 h \citep{walsh2018rubble}.
However, this may not directly apply to 2016 HO$_3$. Despite its rapid rotation period of only 0.5 h, its size is significantly smaller than the range for which the spin barrier is typically relevant (>200 m).
Therefore, whether the small superfast rotator 2016 HO$_3$ is a rubble pile structure remains uncertain.

Theoretical and numerical methods have been developed to analyze the structure stability of asteroids.
\citet{holsapple2001equilibrium,holsapple2004equilibrium} assumed linear elasticity, an ideal ellipsoidal shape, and a stress-free initial state for the asteroid. From these assumptions, he then derived the internal stress field within the asteroid body. Based on such a stress field, it is convenient to establish relations for the maximum spin allowable as a function of the asteroid's internal density, its ellipsoidal shape and the angle of friction, by evaluating the stress field against the Mohr-Coulomb or Drucker-Prager failure model.
Since this theoretical technique required a simplified shape to calculate the stress field and used averaged stress over the whole volume, it discarded information about the failure mode of non-spherical asteroids. 
\citet{hirabayashi2014stress} developed a finite element model suitable for irregular asteroids with uniform material distribution, and show that the center of the asteroid body should fail prior to the surface if with a uniform strength.
However, this approach based on continuum mechanics can not not accurately capture the discrete nature of actual rubble pile asteroids.
Therefore, the Discrete Element Method (DEM) has become a popular alternative for studying the structural stability of granular aggregates.
For instance, \citet{zhang2022inferring} used DEM to simulate the structural history of top-shaped rubble pile asteroid Bennu and estimated an interior strength of only 1.3 Pa.
A similar procedure has been applied to asteroid Didymos, another top-shaped body, to analyze its creep stability ahead of the arrival of the DART mission \citep{zhang2018rotational,zhang2021creep}.
However, these studies focus on near-spherical asteroids of several hundred meters in size, and in these DEM simulations, only spherical constituent particles are used, neglecting irregular boulders that are abundantly present on actual rubble pile asteroids.
2016 HO$_3$, on the other hand, is a highly elongated shaped asteroid with size ranging between 36 to 60 m.
Therefore, it is necessary to conduct a specific analysis for this unique body.

In this paper, we employ the latest shape models and advanced numerical techniques to infer the strength required for 2016 HO$_3$ to maintain structural stability, as well as to infer its possible structure. 
Our analysis focuses on determining the minimum stable strength for typical rubble pile structures. 
By comparing these calculated strength thresholds with existing observational data of asteroid strengths, we aim to ascertain whether 2016 HO$_3$ could be a rubble pile structure. 
That is, if the critical strength is greater than the usual strength of rubble pile asteroids, then the likelihood of 2016 HO$_3$ being a rubble pile structure is reduced, suggesting it might instead be a monolithic structure.
Additionally, we explore the possibility of a loose regolith layer existing on its surface.
This study is expected to provide valuable insights for future missions to 2016 HO$_3$, particularly in designing sampling mechanisms. 
Furthermore, our findings will also enhances our broader knowledge of small superfast rotators and the rotational failure behavior of small bodies.

The paper is organized as follows. 
We first introduce our numerical modeling method and simulation setup in Section \ref{sec:methods}. Note that a "glued-sphere" aggregate method is introduced in this section to handle irregular boulders in DEM simulations.
We then present our results for different possible structures of 2016 HO$_3$ and estimate the critical strength to maintain its structural integrity under its current spin rate in Section \ref{sec:results}.
With the goal of finding out possible structure for 2016 HO$_3$, Section \ref{sec:discussion} compares the measured critical strengths of different structures with the actual measurements of other asteroids.
In Section \ref{sec:conclusion} we summarizes the main findings of this work.

\section{Methods}
\label{sec:methods}

\subsection{Soft-Sphere Discrete Element Method}
\label{sec:ssdem}
In our study, 2016 HO$_3$ with a rubble-pile asteroid is modeled as a grouping of different sized rocks brought together under the influence of gravity. 
Within this rubble pile, each particle is subjected to a variety of interactions with their proximal environment, incorporating both short-range forces—such as mechanical contacts and van der Waals forces—and long-range forces, notably gravitational forces.
To effectively simulate and resolve these complex interactions, we use the three-dimensional N-body numerical code DEMBody\footnote{DEMBody code is available at \href{https://bin-cheng-thu.github.io/dembody-code/}{https://bin-cheng-thu.github.io/dembody-code/}}, which encompasses an implementation of the Soft-Sphere Discrete Element Method (SSDEM) to handle mechanical contacts within constituent rocks \citep{cheng2018collision,cheng2019numerical,cheng2020reconstructing}.

In this code, the interaction between any two grains is driven by the Hertzian normal force and Mindlin–Deresiewicz tangential force \citep{somfai2005elastic}. 
These forces, analogous to the composition of nonlinear springs, viscous dashpots, and frictional sliders, are described by the following equation \eqref{eqa:contact_model}:
\begin{equation}\label{eqa:contact_model}
\begin{array}{*{20}{c}}
{F_{n} = \frac{4}{3}\sqrt {{r_{\mathrm{eff}}}} E_{\mathrm{eff}}^*\delta _{n}^{3/2} - {\gamma _n}{u_n}}\\
{F_{s} = \min \left( {\mu \left| {F_{n}} \right|,8\sqrt {{r_{\mathrm{eff}}}{\delta _{n}}} G_{\mathrm{eff}}^*\delta _{s} - {\gamma _s}{u_s}} \right)}
\end{array},
\end{equation}
Variables ${\delta_{n}}$ and ${\delta_{s}}$ (${u_n}$ and ${u_s}$) denote the normal and tangential mutual displacements (velocities) between the two interacting bodies, respectively. 
Dashpot parameters coefficients, ${\gamma_n}$ and ${\gamma_s}$, correlate with restitution coefficients ${\epsilon_n}$ and ${\epsilon_s}$ as deduced by \cite{wada2006numerical}. 
In this work, ${\epsilon_n}$ and ${\epsilon_s}$ were empirically set as ${\epsilon_n=\epsilon_s=0.55}$, a reasonable value previously adopted in asteroid regolith simulations by \cite{cheng2019numerical}.
The friction coefficient, ${\mu}$, mimics the Mohr-Coulomb sliding criterion, thereby regulating the shear strength of granular materials. 
Furthermore, the effective radius ${r_{\rm{eff}}}$, effective Young's modulus ${E^{*}_{\rm{eff}}}$, and effective shear modulus ${G^{*}_{\rm{eff}}}$ are derived from the harmonic mean of properties such as radius ${r}$, modified Young's modulus ${E^*}$, and modified shear modulus ${G^*}$ of the interacting bodies. 
The modified modulus ${E^*}$ and ${G^{*}}$ were extracted from Young's modulus ${E}$ and Poisson's ratio ${v}$ using ${E^*=E/(1-v^2)}$ and ${G^{*}=E/\left[ {2\left( {1 + v} \right)\left( {2 - v} \right)} \right]}$.

Further, the non-spherical nature of actual grains creates resistance to rolling and twisting motions. 
Consequently, a spring-dashpot rotational model displaying both elastic and plastic characteristics, as introduced by \cite{jiang2015novel}, was incorporated into DEMBody.  
This model integrates rolling and twisting forces into the following equation \eqref{eqa:rolling_model}:
\begin{equation}\label{eqa:rolling_model}
\begin{array}{*{20}{c}}
{{M_r} = \min \left( {0.525\left| {{F_n}} \right|\beta {r_{\rm{eff}}},0.25{{\left( {\beta {r_{\rm{eff}}}} \right)}^2}\left[ {{k_n}{\delta _r} - {\gamma _n}{\omega _r}} \right]} \right)}\\
{{M_t} = \min \left( {0.65\mu \left| {{F_n}} \right|\beta {r_{\rm{eff}}},0.5{{\left( {\beta {r_{\rm{eff}}}} \right)}^2}\left[ {{k_s}{\delta _t} - {\gamma _s}{\omega _t}} \right]} \right)}
\end{array},
\end{equation}
In this equation, ${\delta_r}$ and ${\delta_t}$ (${\omega_r}$ and ${\omega_t}$) represent the total relative rolling and twisting angular displacements (velocities) between two interacting grains, respectively. 
The spring stiffness parameters ${k_n}$ and ${k_s}$ can be computed from the material properties as ${k_n=\frac{4}{3}\sqrt {{r_{\rm{eff}}}{\delta {n}}} E_{\rm{eff}}^*}$ and ${k_s=8\sqrt {{r_{\rm{eff}}}{\delta {n}}} G_{\rm{eff}}^*}$.
The shape parameter ${\beta}$ serves as a statistical measurement of real particle non-sphericity, and its incorporation allows for the consideration of the effects of irregular particle shapes on the mechanical behavior of granular materials. 
The combination of the above two physical parameters, ${\mu}$ and ${\beta}$, effectively replicates the micro- and macro-mechanical behaviors of cohesionless granular materials, offering various internal friction angles $\phi$ within a realistic range as validated in \cite{jiang2015novel} and \cite{cheng2023measuring}.

For small rubble piles residing in low-gravity, ultra-high-vacuum environments, the role of the van der Waals force between constituent grains becomes prominent \citep{scheeres2010scaling}. 
As such, a simple dry cohesion model inspired by the granular bridge concept proposed by \cite{sanchez2014strength} was used to imitate the weak inter-particle forces within asteroid regolith \citep{zhang2018rotational,cheng2019numerical}.
\begin{equation}
{F_c=c(\beta r_{\rm{eff}})^2},
\end{equation}
Here, $(\beta r_{\rm{eff}})^2$ symbolizes the contact area between two interacting particles. 
The microscopic tensile strength ${c}$ governs the macroscopic cohesion strength of granular materials, ranging from a few Pa to several tens of Pa as deduced from previous observations.
Note that the cohesive force is considered only for contacting particles, while disregarded once the particles are separated.

\begin{table*}
	\centering
	\caption{Summary of Simulation Parameters (First Column to Third Column) and Cohesion Strength (Fourth Column to Fifth Column) and Critical Spin Period (Sixth Column) and Minimum Cohesion Strength Required for 2016 HO$_{3}$ (Last Column).}
    \begin{threeparttable}
	\label{tab:parameters}
	\begin{tabular}{ccccccc}
		\hline
		Asteroid shape & Rubble-pile structure & Grain size (m) & $c$ (Pa) & $C$ (Pa) & $T_\mathrm{crit}$ (h) & $C_\mathrm{HO3}$ (Pa)\\
		\hline
		~ & \multirow{10}*{Fine-grained} & \multirow{10}*{0.2-0.7} & 0 & 0 & 4.588 & \multirow{10}*{8.34}\\
        ~ & ~ & ~ & 1 & 0.006 & 4.324 & ~\\
        ~ & ~ & ~ & 5 & 0.032 & 3.768 & ~\\
        ~ & ~ & ~ & 42.5 & 0.271 & 2.182 & ~\\
        Shape I & ~ & ~ & 72.5 & 0.462 & 1.745 & ~\\
        ($59 \times 28 \times 28$ m) & ~ & ~ & 317.5 & 2.025 & 0.970 & ~\\
        ~ & ~ & ~ & 500.0 & 3.189 & 0.720 & ~\\
        ~ & ~ & ~ & 702.5 & 4.481 & 0.582 & ~\\
        ~ & ~ & ~ & 1350.0 & 8.610 & 0.436 & ~\\
        ~ & ~ & ~ & 2000.0 & 12.756 & 0.364 & ~\\
		\hline
		~ & \multirow{8}*{Fine-grained} & \multirow{8}*{0.2-0.7} & 1 & 0.006 & 4.551 & \multirow{8}*{13.06}\\
        ~ & ~ & ~ & 5 & 0.032 & 3.920 & ~\\
        ~ & ~ & ~ & 42.5 & 0.271 & 2.463 & ~\\
        Shape II & ~ & ~ & 72.5 & 0.462 & 2.006 & ~\\
        ($66 \times 32 \times 22$ m) & ~ & ~ & 317.5 & 2.025 & 1.097 & ~\\
        ~ & ~ & ~ & 702.5 & 4.481 & 0.771 & ~\\
        ~ & ~ & ~ & 1350.0 & 8.610 & 0.567 & ~\\
        ~ & ~ & ~ & 3000.0 & 19.134 & 0.391 & ~\\
        \hline
		~ & \multirow{8}*{Fine-grained} & \multirow{8}*{0.2-0.7} & 1 & 0.006 & 5.878 & \multirow{8}*{20.31}\\
        ~ & ~ & ~ & 5 & 0.032 & 5.023 & ~\\
        ~ & ~ & ~ & 42.5 & 0.271 & 3.119 & ~\\
        Shape III & ~ & ~ & 72.5 & 0.462 & 2.619 & ~\\
        ($89 \times 27 \times 19$ m) & ~ & ~ & 317.5 & 2.025 & 1.283 & ~\\
        ~ & ~ & ~ & 702.5 & 4.481 & 0.906 & ~\\
        ~ & ~ & ~ & 1350.0 & 8.610 & 0.650 & ~\\
        ~ & ~ & ~ & 4000.0 & 25.511 & 0.430 & ~\\
        \hline
		~ & \multirow{8}*{Fine-grained} & \multirow{8}*{0.2-1.0} & 1 & 0.006 & 4.304 & \multirow{8}*{7.03}\\
        ~ & ~ & ~ & 5 & 0.032 & 2.991 & ~\\
        ~ & ~ & ~ & 42.5 & 0.271 & 2.001 & ~\\
        Shape I & ~ & ~ & 72.5 & 0.462 & 1.673 & ~\\
        ($59 \times 28 \times 28$ m) & ~ & ~ & 317.5 & 2.025 & 0.876 & ~\\
        ~ & ~ & ~ & 702.5 & 4.481 & 0.605 & ~\\
        ~ & ~ & ~ & 1350.0 & 8.610 & 0.436 & ~\\
        ~ & ~ & ~ & 2000.0 & 12.756 & 0.361 & ~\\
        \hline
		~ & \multirow{8}*{Fine-grained} & \multirow{8}*{0.2-2.0} & 1 & 0.006 & 3.956 & \multirow{8}*{7.50}\\
        ~ & ~ & ~ & 5 & 0.032 & 3.400 & ~\\
        ~ & ~ & ~ & 42.5 & 0.271 & 1.950 & ~\\
        Shape I & ~ & ~ & 72.5 & 0.462 & 1.636 & ~\\
        ($59 \times 28 \times 28$ m) & ~ & ~ & 317.5 & 2.025 & 0.855 & ~\\
        ~ & ~ & ~ & 702.5 & 4.481 & 0.585 & ~\\
        ~ & ~ & ~ & 1350.0 & 8.610 & 0.436 & ~\\
        ~ & ~ & ~ & 2000.0 & 12.756 & 0.350 & ~\\
        \hline
		~ & \multirow{8}*{Surface layer\tnote{a}} & \multirow{8}*{0.2-0.7} & 1 & 0.006 & 3.614 & \multirow{8}*{1.26}\\
        ~ & ~ & ~ & 5 & 0.032 & 3.071 & ~\\
        ~ & ~ & ~ & 10 & 0.064 & 1.828 & ~\\
        Shape I & ~ & ~ & 50 & 0.319 & 0.905 & ~\\
        ($59 \times 28 \times 28$ m) & ~ & ~ & 100 & 0.638 & 0.664 & ~\\
        ~ & ~ & ~ & 150 & 0.957 & 0.562 & ~\\
        ~ & ~ & ~ & 200 & 1.276 & 0.498 & ~\\
        ~ & ~ & ~ & 317.5 & 2.025 & 0.425 & ~\\
        \hline
		~ & \multirow{8}*{Boulder scatter\tnote{b}} & \multirow{8}*{0.2-10.0} & 1 & 0.006 & 4.385 & \multirow{8}*{8.18}\\
        ~ & ~ & ~ & 5 & 0.032 & 3.949 & ~\\
        ~ & ~ & ~ & 42.5 & 0.271 & 2.544 & ~\\
        Shape I & ~ & ~ & 72.5 & 0.462 & 1.532 & ~\\
        ($59 \times 28 \times 28$ m) & ~ & ~ & 317.5 & 2.025 & 1.056 & ~\\
        ~ & ~ & ~ & 702.5 & 4.481 & 0.641 & ~\\
        ~ & ~ & ~ & 1350.0 & 8.610 & 0.453 & ~\\
        ~ & ~ & ~ & 2000.0 & 25.51 & 0.373 & ~\\
        \hline
	\end{tabular}
    \begin{tablenotes}
        \item[a] $C$ in the surface layer model represent the strength of the uppermost 2 m loose layer.
        \item[b] $C$ in the boulder scatter model represent the strength of the interstitial regolith amongst the boulders.
    \end{tablenotes}

    \end{threeparttable}
\end{table*}

\subsection{Micro-macro transition of granular quantities}
\label{sec:micro2macro}
As a discrete system, the microscopic physical information of individual particles in granular materials is insufficient to reflect the dynamic behavior of the entire system, thus necessitating the acquisition of macroscopic properties.
This has previously been done by employing a homogenization and averaging method within Representative Volume Elements (RVEs), such as in \cite{zhang2018rotational}.
However, through such homogenization, the macroscopic properties within the same RVE are artificially set to be identical, failing to adequately characterize the variations at the particle scale.
This limitation is distinctly discernible as shown by the artificial grid distribution of the stress field in the paper by \cite{zhang2018rotational}.

\begin{figure}
    \centering
	\includegraphics[clip,trim=0mm 0mm 0mm 0mm,width=0.9\linewidth]{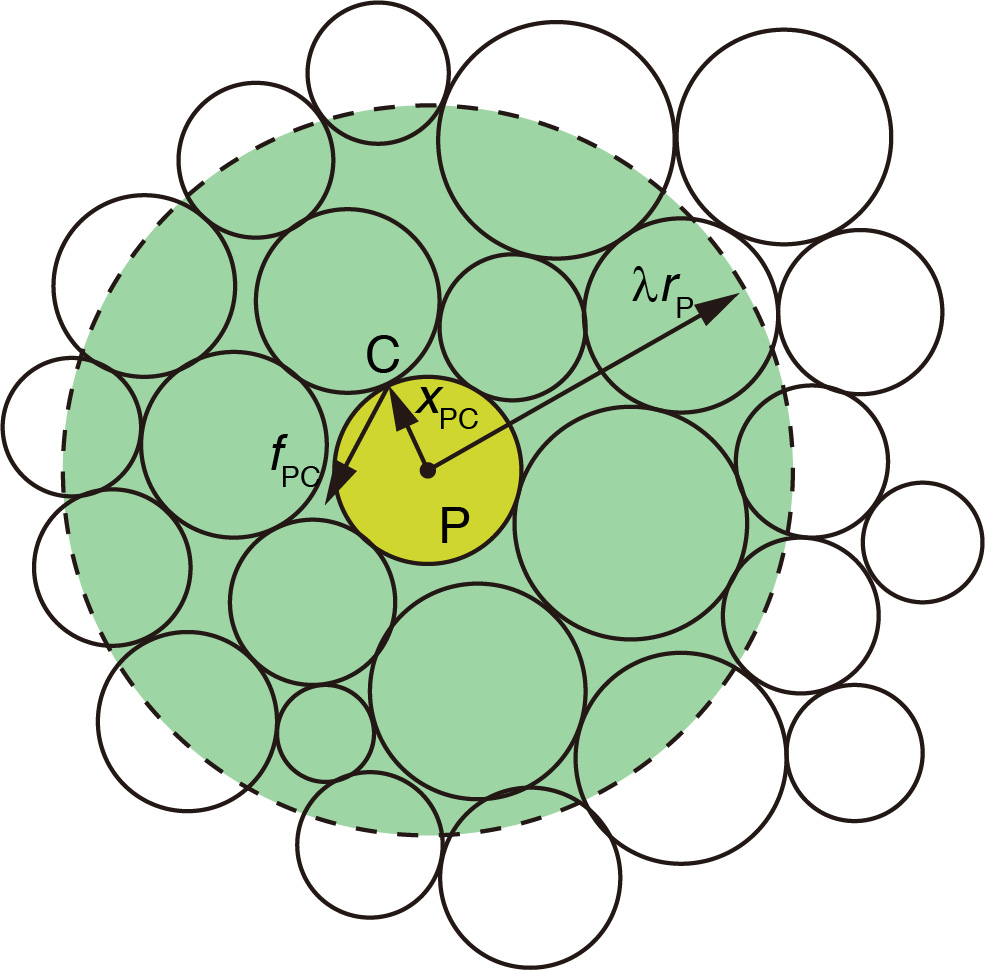}
    \caption{The coarse-grained domain (the green region) with radius of $\lambda r_p$ for a particle with radius $r_p$ in a granular assembly. This particle interacts with surrounding particles through the branch vector $\boldsymbol{x}_{PC}$, connecting the particle center with the contact point, and the corresponding contact force $\boldsymbol{f}_{PC}$.}
    \label{fig:domain}
\end{figure}

Here we propose a spatial coarse-grained approach to extract the macroscopic properties while simultaneously reflecting variations at the particle scale.
In the microscopic scale, the relevant quantities are local and attached to the particle or to the contact scale: the contact force, acting at the contact area between adjoining particles, and the relative motion between particles.
Therefore, obtaining macroscopic properties requires to expand the definition domain to a larger scale where the assumptions of the continuum hold.
This scale is referred to as the coarse-grained domain.
As shown in Fig.~\ref{fig:domain}, for each particle with radius $r_p$, a circle of radius $\lambda r_p$ centered at the particle is defined as its coarse-grained domain, and all the particles in this circle are considered to be its influenced neighbors.
We choose the size ratio $\lambda$ to be 5, which has been verified to not only eliminate fluctuations at the microscopic scale but also reflect the variation within the whole computational domain.

\begin{figure*}
	\includegraphics[clip,trim=0mm 0mm 0mm 0mm,width=0.8\textwidth]{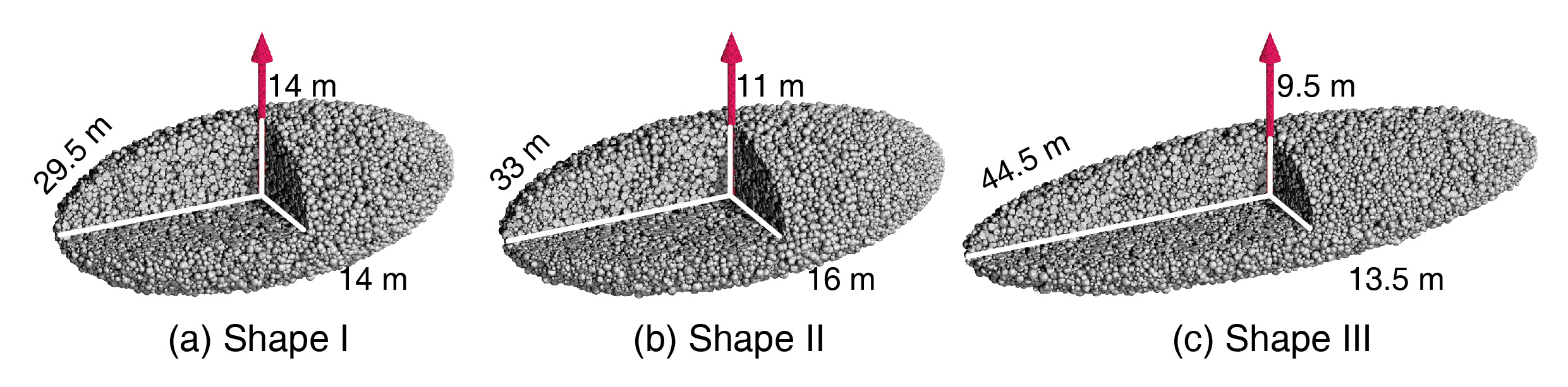}
    \caption{Different shape models of 2016 HO$_{3}$ used in this work.}
    \label{fig:morphology}
\end{figure*}

\subsubsection{Porosity}
Based on the coarse-grained domain defined above, the packing fraction $\eta_p$ of a particle can be calculated as the ratio of the cumulative volume of all particles within the domain to the volume of the domain itself as
\begin{equation}\label{eqa:packingratio}
\eta_p = \frac{\sum_{k=1}^{N_p} V_k}{\frac{4}{3} \pi (\lambda r_p)^3},
\end{equation}
where $N_p$ is the total number of particles inside the domain, and $V_k$ is the volume of individual particles inside the domain.
The average packing fraction of the entire assembly with $N$ particles is then given by 
\begin{equation}
\eta_A = \frac{1}{N} \sum_{k=1}^{N}\eta_k,
\end{equation}
Note that for particles proximate to the surface of the granular assembly, their coarse-grained domains encompass a substantial extent of void space, thereby leading to a considerable underestimation of their packing fractions.

\subsubsection{Stress}
For granular assemblies that are in static equilibrium state, the gravity forces and inertial effects can be omitted, threfore, the averaged Cauchy stress tensor for a particle is given according to Cauchy's first law of motion as \citep{nicot2013definition}
\begin{equation}
   \label{eqa:stresstensor}
   \boldsymbol{\overline {\sigma}} = \frac{1}{V} \sum_{P \in V}^{} \sum_{C \in P}^{} \boldsymbol{x}_{PC} \otimes \boldsymbol{f}_{PC}.
\end{equation}
where $V$ is the volume of the coarse-grained domain with radius of $\lambda r_p$. 
The equation represents the summation of the dyadic product of the branch vector $\boldsymbol{x}_{PC}$ and the contact force $\boldsymbol{f}_{PC}$ over all the contacts `\textit{C}' between two adjoining particles `\textit{P}' inside the corresponding domain `\textit{V}'.
Tension is negative and compression positive.

With the derived principal stresses of the average stress tensor $\{\overline {\sigma}_{1} \geq \overline {\sigma}_{2} \geq \overline {\sigma}_{3}\}$, its first invariant $\overline I_{1}$ and the deviatoric stress $\overline J_{2}$ can then be calculated by,
\begin{equation}
   \begin{array}{*{20}{c}}
   \overline I_{1} = \overline {\sigma}_{1} + \overline {\sigma}_{2} + \overline {\sigma}_{3},\\
   \overline J_{2} = [(\overline {\sigma}_{1} - \overline {\sigma}_{2})^2 + (\overline {\sigma}_{2} - \overline {\sigma}_{3})^2 + (\overline {\sigma}_{3} - \overline {\sigma}_{1})^2]/6,
   \end{array}
\end{equation}
Here $\overline I_{1}$ measures the pressure acting on the central particle, and we
use it to evaluate the distribution of the dynamical internal pressure within the simulated rubble pile.

\subsubsection{Strain and non-affine stain}
Next, we consider rheology of the granular particles during its spin-up failure.
This will help to provide information about the transition between solid and fluid states in granular materials.
Affine deformation refers to uniform displacement of particles which preserves their relative positioning, thus typically associated with the elastic response of the material.
While non-affine deformation disrupts the initial geometrical relationships between particles, often leading to irreversible, plastic deformations that indicate the yield and failure of granular materials. 

The affine and non-affine components are computed using the approach from \cite{falk1998dynamics}, outlined briefly in what follows.
For a specific particle with its central position of $\boldsymbol{x}_{0}(t)$, the neighboring particles within its coarse-grained domain have central positions of $\boldsymbol{x}_{k}(t)$, where $k$ can be 1 to $N_{\rm{neigh}}$ and $N_{\rm{neigh}}$ is the number of neighboring particles of the central particle.
Then, the branch vector of one neighboring particle relative to the central one is given by $\boldsymbol{x}_{k}(t) - \boldsymbol{x}_{0}(t)$ at time $t$.
Assuming only elastic (affine) strains are present, we can predict the new branch vector at time $t+\delta t$ as
\begin{equation}\label{eqa:affinepart}
    \boldsymbol{A}(t) [\boldsymbol{x}_{k}(t) - \boldsymbol{x}_{0}(t)]
\end{equation}
where the affine matrix $\boldsymbol{A}(t)$ is a $3\times 3$ matrix.
The difference between an actual deformation and the affine part is defined as the non-affine deformation.
It can be measured by a quantity $\mathcal{D}^2$ as the mean-square difference between the actual branch vector $\boldsymbol{x}_{k}(t+\delta t) - \boldsymbol{x}_{0}(t+\delta t)$ and the predicted one defined by Eq.~\ref{eqa:affinepart}, that is
\begin{equation}\label{eqa:nonaffinepart}
    \mathcal{D}^2 = \sum_{k=1}^{N_\mathrm{neigh}} \left \|  \boldsymbol{x}_{k}(t+\delta t) - \boldsymbol{x}_{0}(t+\delta t) - {\boldsymbol{A}(t) [\boldsymbol{x}_{k}(t) - \boldsymbol{x}_{0}(t)]} \right \|^2
\end{equation}
Then we minimize $\mathcal{D}^2$ to get the minimum $\mathcal{D}^2_{\mathrm{min}}$ by optimizing the affine matrix $\boldsymbol{A}(t)$.
The scaled non-affine stain $\mathcal{D}_{\mathrm{min}} / r$ indicates the local non-affine component of strain in the vicinity of each particle, which provide a quantitative understanding of the granular assembly's rheology and failure behavior. 

\subsection{Simulation setup}
\subsubsection{Possible shapes of 2016 HO$_{3}$}
Based on the limited light curve data of 2016 HO$_{3}$, three possible triaxial shape models were derived for this asteroid in \cite{li2021shape} as shown in Fig.~\ref{fig:morphology}.
We denote the 59 m $\times$ 28 m $\times$ 28 m ellipsoid as shape I; the 66 m $\times$ 32 m $\times$ 32 m ellipsoid as shape II; the 89 m $\times$ 27 m $\times$ 19 m ellipsoid as shape III, with the $c$-axis further shortened and in turn the $a$-axis much longer.
Note that the asteroid rotates about the $c$-axis.
In subsequent sections, we employ these three shape models to analyze the structural stability of the asteroid 2016 HO$_{3}$ and infer its possible interior structure.

\subsubsection{2016 HO$_{3}$ rubble-pile models}
2016 HO$_{3}$ is explicitly modelled as a self-gravitating rubble pile in our numerical investigation.
Initially, a granular assembly slightly larger than 2016 HO$_{3}$ is prepared by letting a point cloud of rocks with random velocities but without contacts to collapse together under the self-gravity.
The initial point cloud is set to be with a pre-defined size distribution that follows a a power law characterized by an exponent of -3.0.
After all particles in the aggregate settle down, we use the shape model as described above to pare superfluous particles down to sculpt the assembly into 2016 HO$_{3}$'s likeness.
Then we let this trimmed assembly to settle down once again with the corresponding material parameters for different tests.

\subsubsection{Rubble-pile structures}
\begin{figure}
\centering
	\includegraphics[clip,trim=0mm 0mm 0mm 0mm,width=0.7\linewidth]{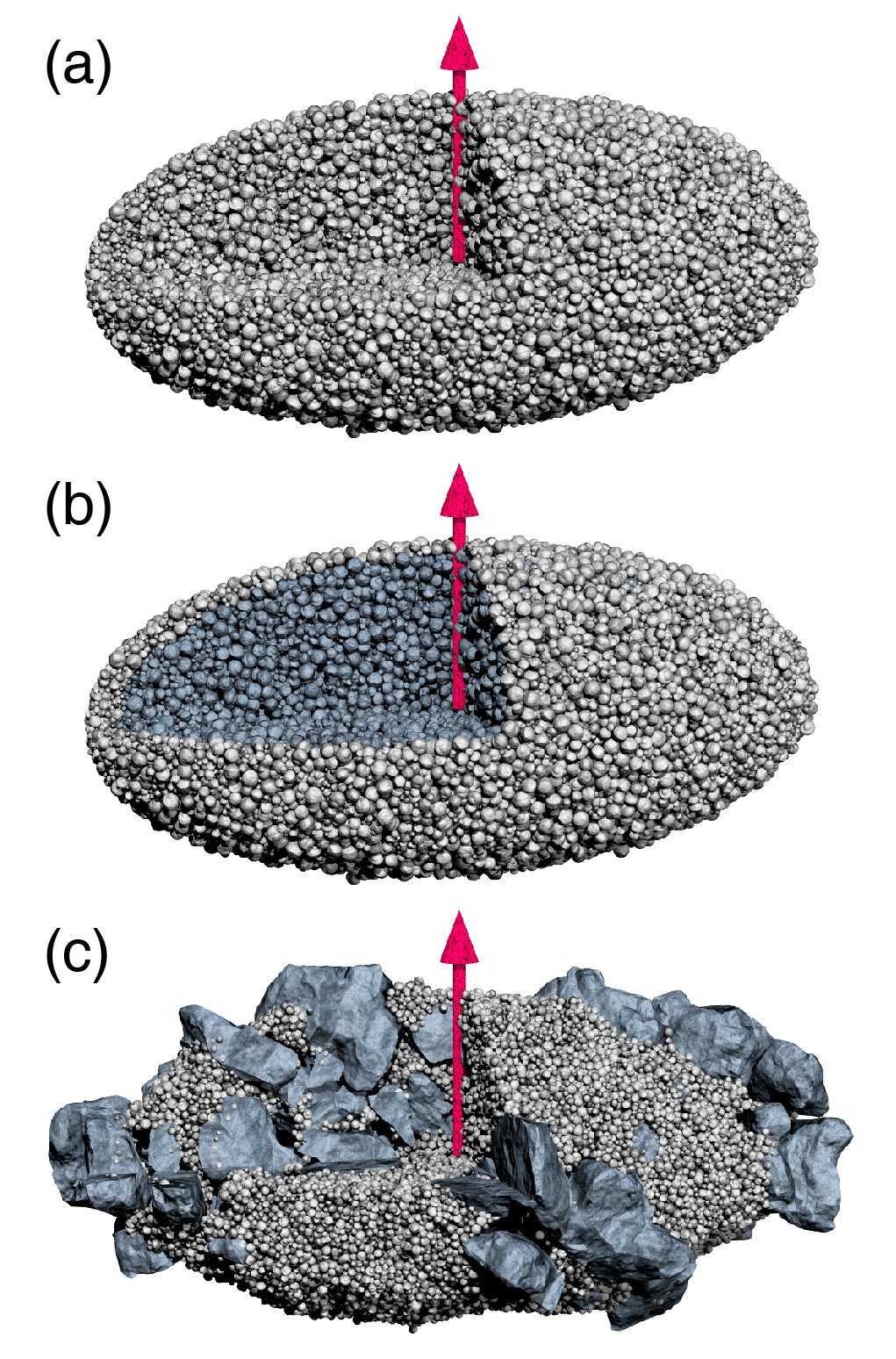}
    \caption{Cut-away schematic of possible rubble-pile structures of 2016 HO$_3$.
    (a), the fine-grained model with dense polydisperse packing configuration and homogeneous macroscopic strength. 
    (b), the surface layer model with a loose regolith surface (grey grains) underlain by a strong core (blue grains).
    (c), the boulder scatter model with highly-irregular boulders (blue polyhedrons) strewn in loose interstitial regolith (grey grains).}
    \label{fig:structureSetting}
\end{figure}

To investigate the possible structure of 2016 HO$_3$, we explore three possible rubble-pile configurations in this study as shown in Fig.~\ref{fig:structureSetting}: 1. a fine-grained model; 2. a surface layer model; and 3. a boulder scatter model.
The fine-grained model represents ordinary granular aggregates with 
dense polydisperse packing configuration and homogeneous macroscopic strength.
We use three different particle size ranges, 0.2-0.7 m, 0.2-1.0 m and 0.2-2.0 m, for the fine-grained model.
The surface layer model is inspired by the layered structure of Ryugu evidenced by the artificial impact experiment.
\cite{arakawa2020artificial} found that there was a much cohesive basement layer under the loose surface of Ryugu.
Therefore, we use a 0.2-0.7 m polydisperse packing rubble pile model, and set its uppermost 2 m layer with various low cohesive strengths, underlain by a strong core with a strength of 1 kPa.
The above two models adopt granular aggregates composed of spherical particles with comparable sizes.
However, previous observations found asteroids are strewn with an abundance of irregular boulders, some of which attain sizes as large as 20\% of the asteroid's overall size, such as the Otohime boulder found on asteroid Ryugu \citep{watanabe2019hayabusa2}.
Therefore, we arrange arbitrary numbers of highly-irregular boulders into the surface and interior of a 0.2-0.7 m polydisperse packing rubble pile model.
The non-spherical boulders are modelled as "glued-sphere" aggregates in DEMBody, i.e., a boulder is consisted of arbitrary numbers of spherical particles that are assembled into desired shape and then fixed together to behave as a unit.
The interstitial regolith are modelled as small spherical particles with various low cohesive strengths.

\subsubsection{Spin-up process using accelerate-relax alternating method}
In contrast to previous methods that directly set the simulated asteroid’s spin period $T$ to a prescribed value as a function of time \citep{zhang2018rotational}, we design a quasi-static spin-up protocol that more closely approximates the actual spin evolution of asteroids under small torque. It is named as Accelerate-Relax Alternating method.

\begin{figure*}
	\includegraphics[clip,trim=0mm 0mm 0mm 0mm,width=0.9\linewidth]{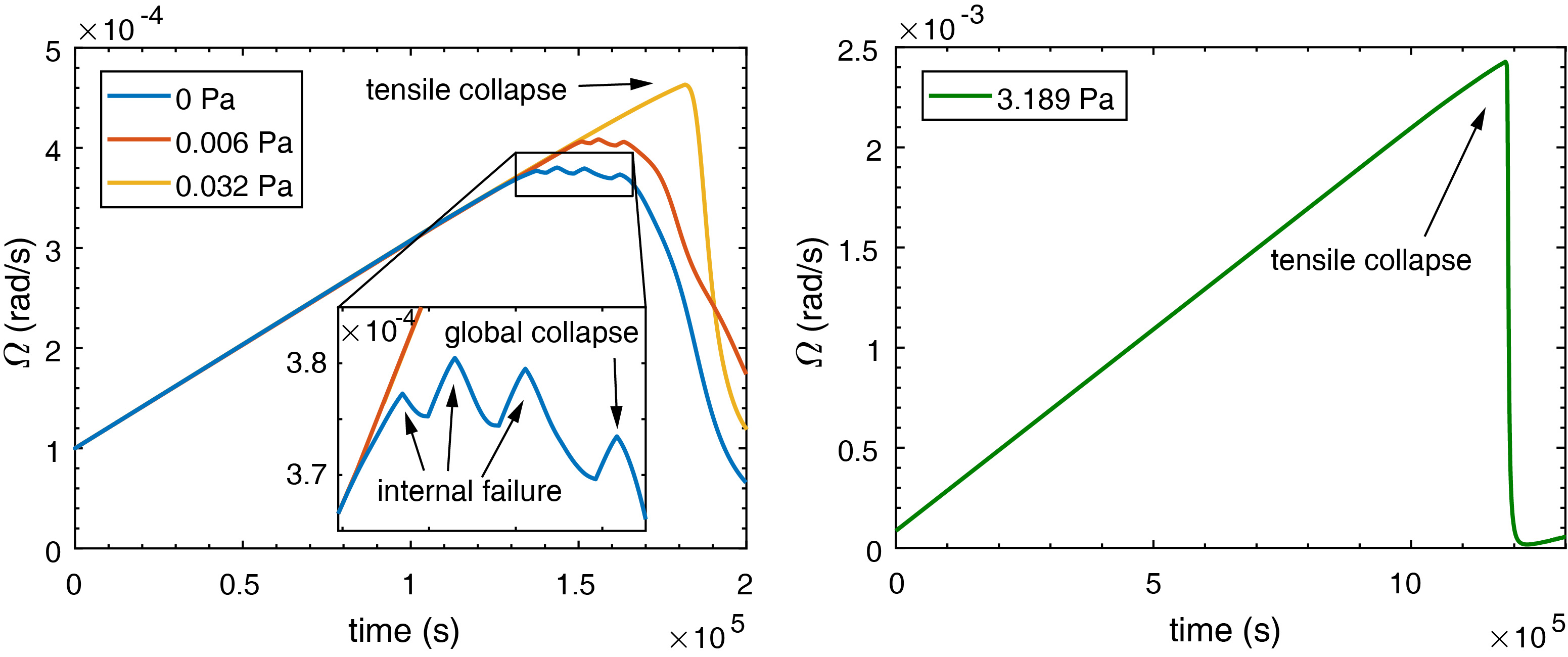}
    \caption{Evolution of spin rate $\Omega$ during spin-up processes for the Shape I rubble-piles with cohesion of 0 Pa, 0.006 Pa, 0.032 Pa and 3.189 Pa, respectively. Initially, its spin rate increases linearly, strictly following the prescribed acceleration path. The deviations from the prescribed path imply structure failures. If the deformation ceases rapidly, then the spin rate approaches stability, indicating that the failure occurred only locally, as shown by the internal failures in the Inset. 
    However, if the spin rate undergoes a rapid decline without tending towards stability, it suggests a global-scale collapse, such as the last turning point at the end of the blue curve.}
    \label{fig:spinratepath}
\end{figure*}

We first let the body settle down at a sufficiently slow spin rate $\Omega_\mathrm{start}$, and then continually apply angular momentum increments at a low constant speed, making the body to spin up accordingly, as shown in Fig.~\ref{fig:spinratepath}.
In this work, we use a spin-up rate of $2 \times 10^{-9}$ $\mathrm{rad/s^2}$, which ensures the structural response of the body to be slow enough to be readily modelled, despite it is still much larger than the typical values due to the YORP torque.
Note that only $c$-axis rotation is considered in this study.
During the spin-up process, we monitor the actual spin rate of the body based on its angular momentum and its moment of inertia.
At first, the spin rate of the simulated body will strictly follow the predefined spin-up
path.
When the measured increase in spin rate after each angular momentum increment is less than the predescribed change by more than 10$\%$, it means that structural failure occurs. 
At this point, the increase in angular momentum is turned off, and the system is allowed to evolve freely under its self-gravity until its spin rate tends to stabilize. 
Then we start to increase the angular momentum again and repeat the above process.
If the spin rate of the system continues to drop significantly and does not tend towards stability, this means the simulated body has collapsed globally, thus there is no need to restart the spin-up process. 
This spin-relax-spin process accurately capture the mechanical response of self-gravitating rubble piles.
The highest spin of the body can reach with various material strengths $C$ is defined as the critical failure spin period $T_\mathrm{crit}$ for the corresponding strengths.
In turn, the minimum strength $C_\mathrm{HO3}$ required to maintain structural stability at 2016 HO$_{3}$'s current spin period (28 min) can be obtained accordingly.

In this work, friction parameters ${\mu}$ and $\beta$ are both maintained to 0.5, which corresponds to a internal friction angle $\phi$ of $\sim 25^\circ$ according to the calibration experiments by \cite{cheng2023measuring}.
And we explore a vast range of $c$ from 1 to 4000 Pa, equivalent to the macroscopic strength $C$ from 0.006 to 25.511 Pa through
\begin{equation}
    C = \frac{c {\beta^2} {N_A} {\eta_A} \tan {\theta}}{4 \pi}
\end{equation}
where $N_A$ is the average coordination number of the entire assembly.
We also investigate the influence of asteroid shapes, grain size distributions and asteroid structures.
All physical properties explored in this paper are summarized in Table \ref{tab:parameters}.

\begin{figure*}
	\includegraphics[clip,trim=0mm 0mm 0mm 0mm,width=0.9\linewidth]{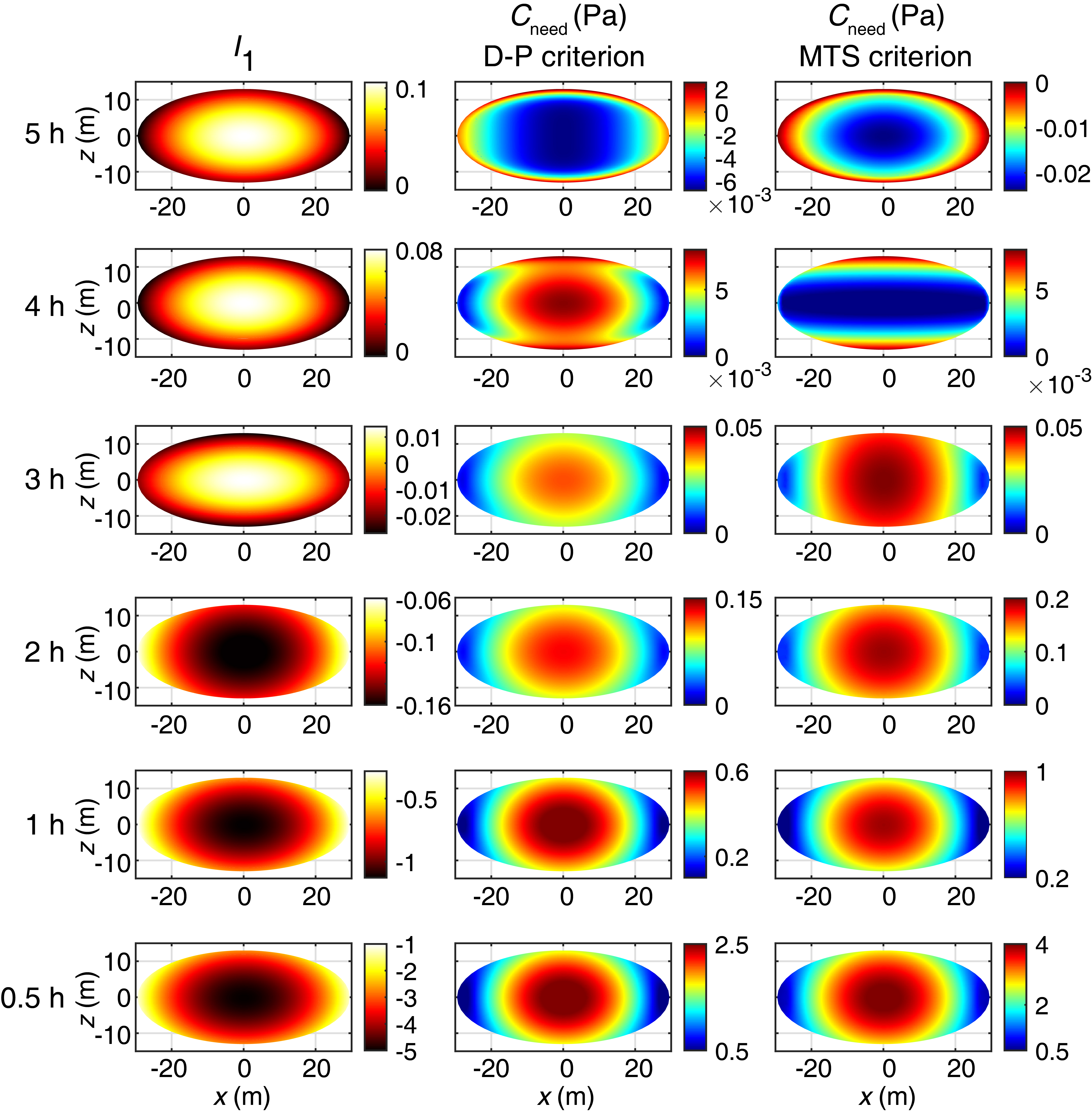}
    \caption{Distribution of $I_1$ (left column), $C_\mathrm{need}$ derived by D-P criterion (middle column) and $C_\mathrm{need}$ derived by MTS criterion (right column) over the cross-section along the $x$ and $z$ axes for the 2016 HO$_{3}$ Shape I model in various spin periods (5 h to 0.5 h). A friction angle of $25^\circ$ and Poisson’s ratio of 0.3 are used for all cases.}
    \label{fig:theoryEvolution}
\end{figure*}

\section{Results}
\label{sec:results}

\begin{figure*}
	\includegraphics[clip,trim=0mm 0mm 0mm 0mm,width=0.9\textwidth]{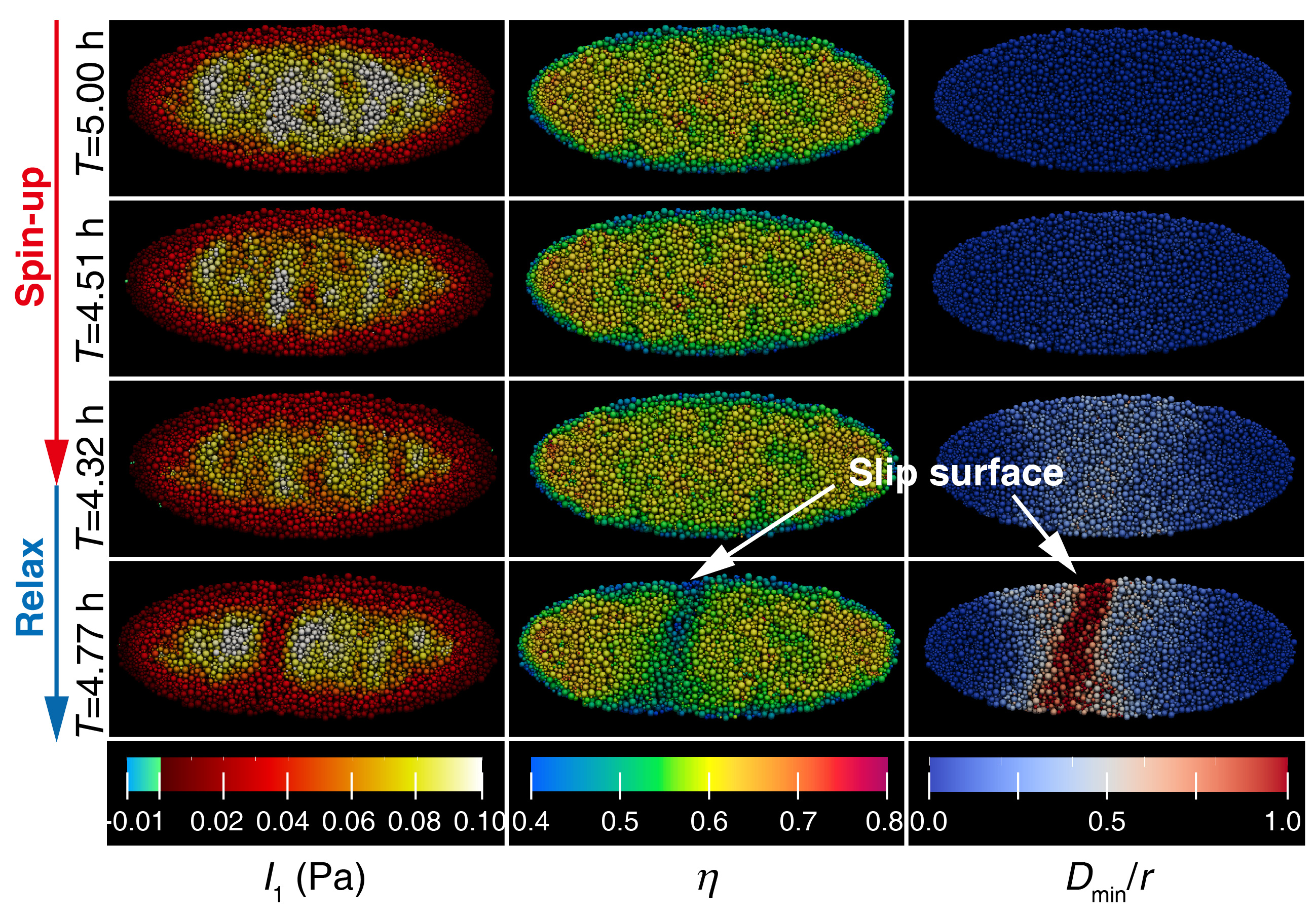}
    \caption{Distribution of the pressure $I_1$, the packing ratio $\eta$ and the scaled non-affine strain $D_\mathrm{min}/r$ at different spin periods over a cross section parallel to the spin axis for Shape I rubble-pile model with cohesive strength of 0.006 Pa. The simulated body is spun up from 5.00 h to 4.32 h, at which time shear failure occurs near its central region. This leads to the body to slow down to 4.77 h due to the relaxation.
    The stress analyses are carried out based on our proposed spatial coarse-grained approach.}
    \label{fig:evolution0.006Pa}
\end{figure*}

\begin{figure*}
	\includegraphics[clip,trim=0mm 0mm 0mm 0mm,width=0.9\textwidth]{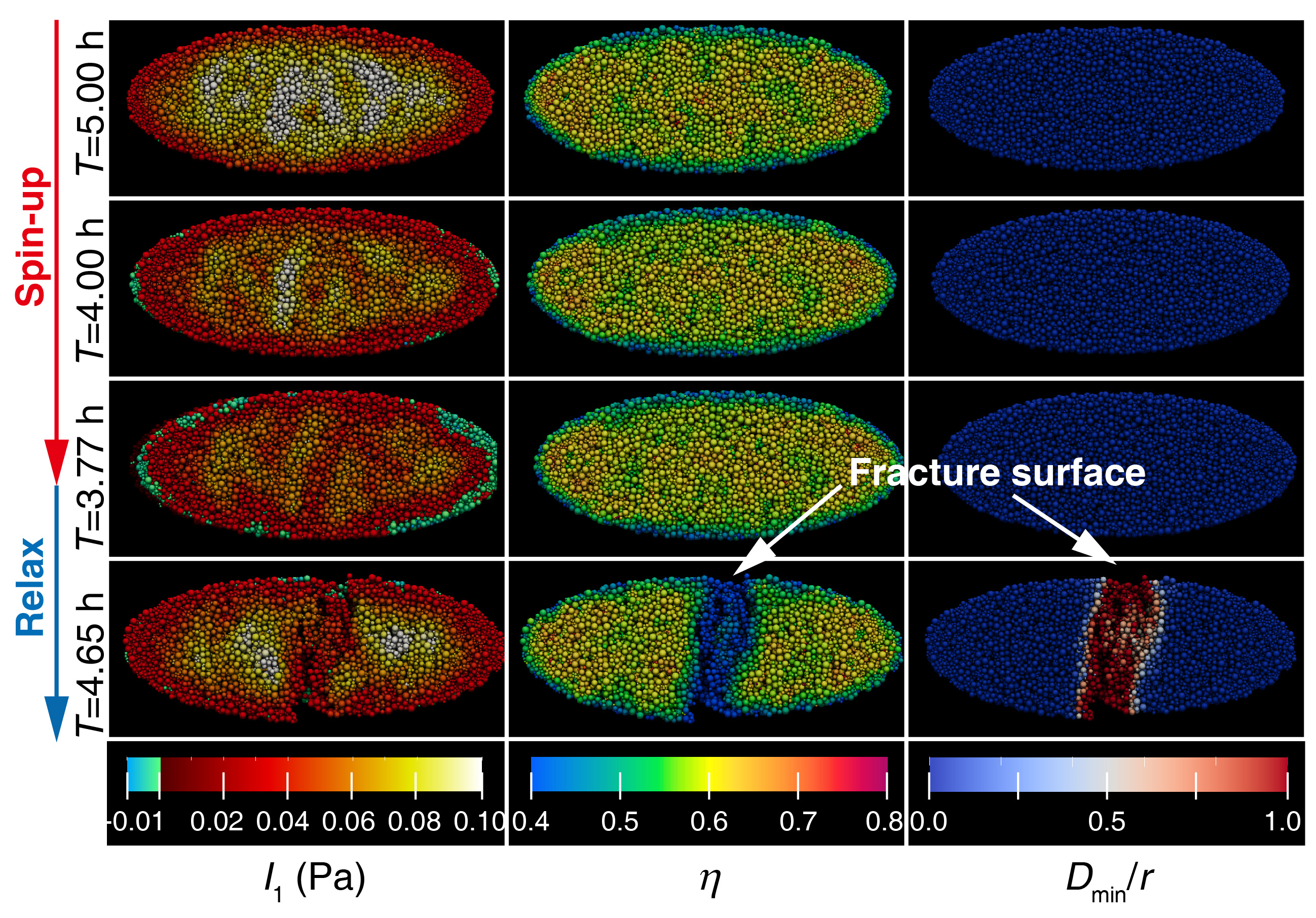}
    \caption{Same as Figure \ref{fig:evolution0.006Pa}, but with a cohesive strength of 0.032 Pa. The simulated body is spun up from 5.00 h to 3.77 h, at which time tensile failure occurs near the middle area. This leads to the body to slow down to 4.65 h due to the relaxation.}
    \label{fig:evolution0.032Pa}
\end{figure*}

\begin{figure*}
	\includegraphics[clip,trim=0mm 0mm 0mm 0mm,width=0.9\textwidth]{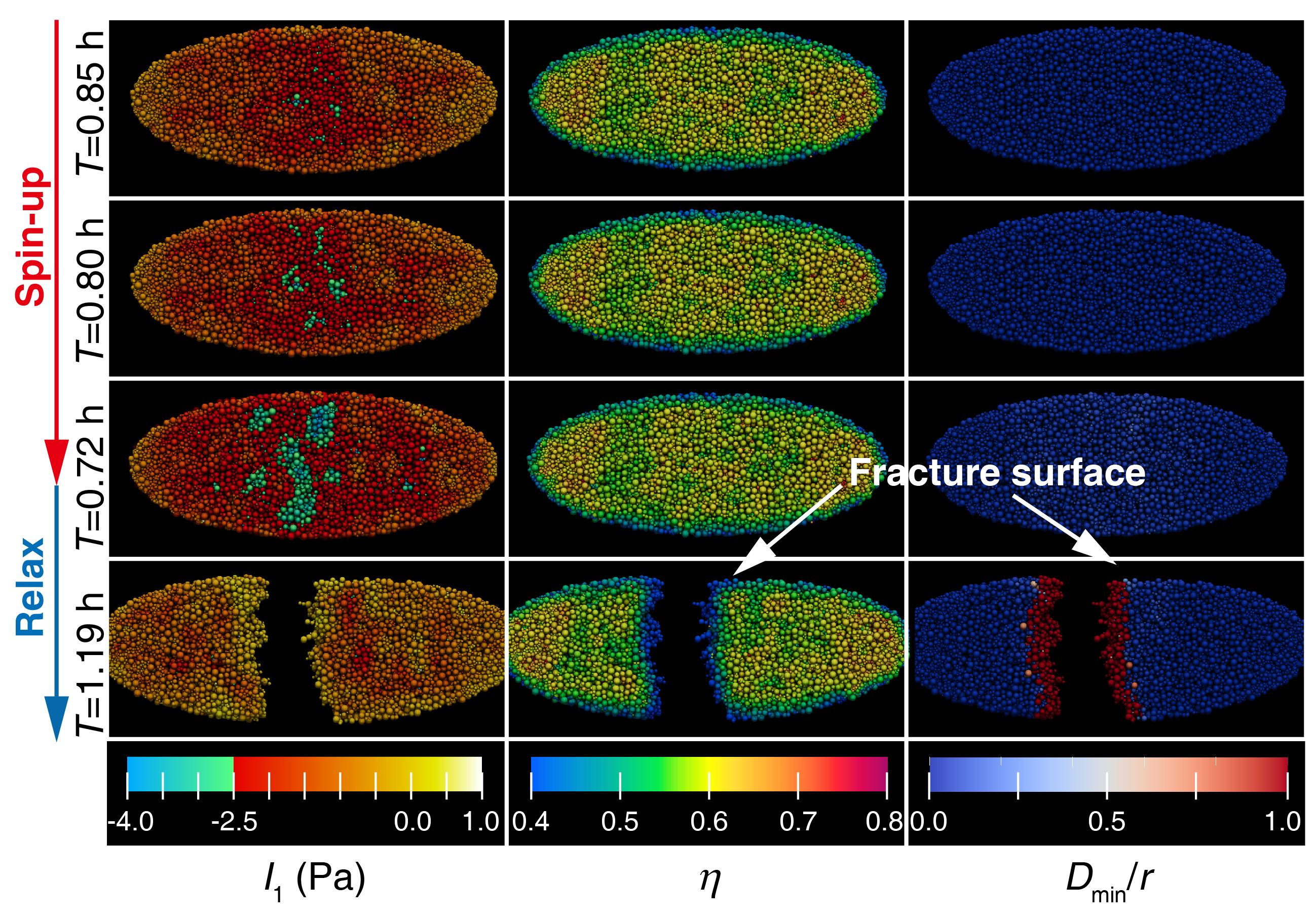}
    \caption{Same as Figure \ref{fig:evolution0.006Pa}, but with a cohesive strength of 3.189 Pa. The simulated body is spun up from 0.85 h to 0.72 h, at which time tensile failure occurs near the middle area. This leads to the body to slow down to 1.19 h due to the relaxation.}
    \label{fig:evolution3.189Pa}
\end{figure*}

\begin{figure}
	\includegraphics[clip,trim=0mm 0mm 0mm 0mm,width=\columnwidth]{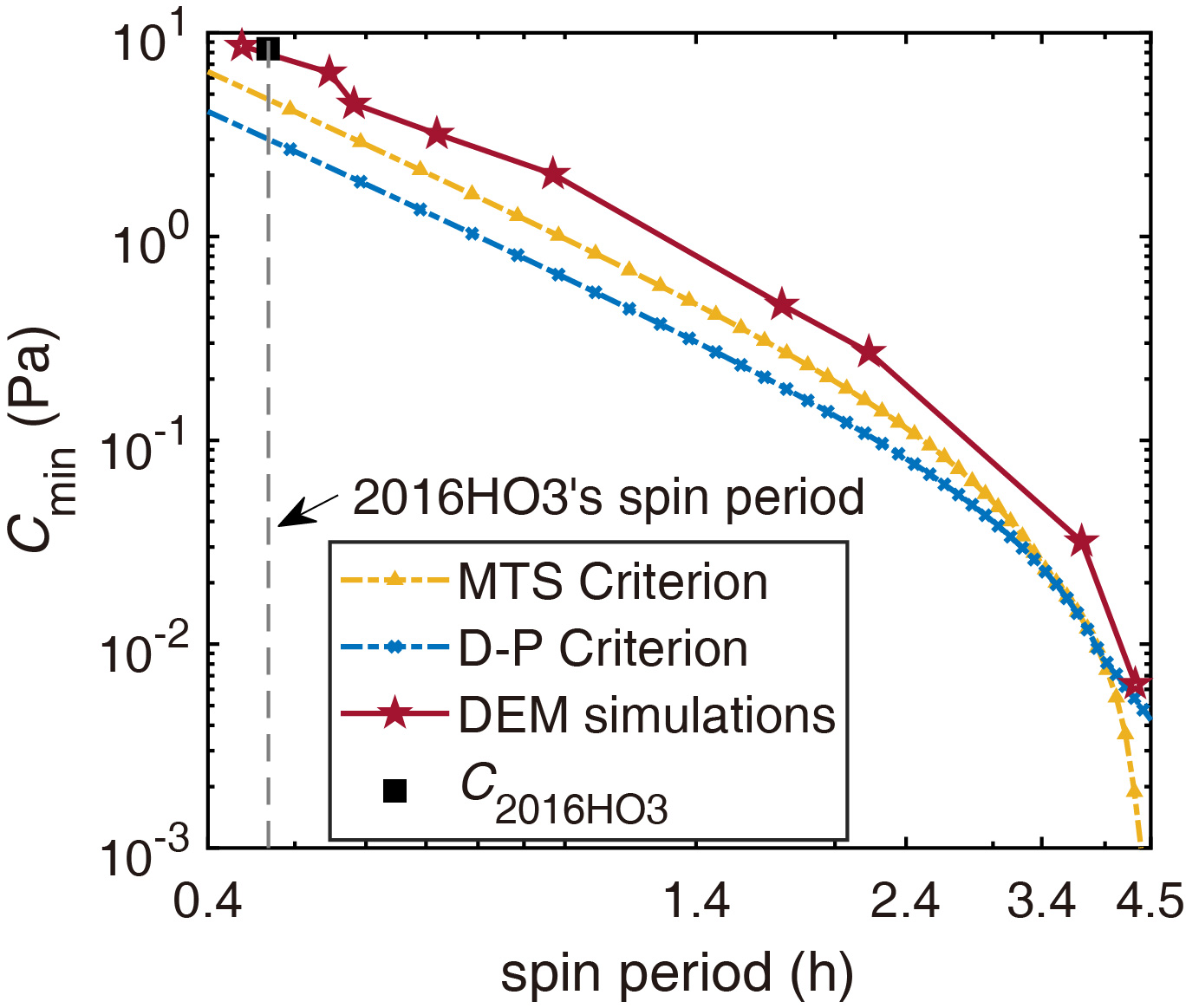}
    \caption{Minimum cohesive strength required for global structural integrity $C_\mathrm{min}$ for Shape I DEM simulations (red line), and theoretical predictions derived by the MTS criterion (yellow line) as well as the D-P criterion (blue line) against spin period.
    The transition from shear failure regime to tensile failure stage found in DEM simulations is consistent with the intersection of the two theoretical curves.
    The black square stands for the minimum cohesion for asteroid 2016 HO$_{3}$ with spin period of 28 min.}
    \label{fig:cmin_model1}
\end{figure}

\subsection{Continuum Mechanics Theory}
Several analytical methods based on continuum mechanics theory have been developed to elucidate the failure mode of a cohesive, spinning rubble-pile.
For example, assuming the body is homogeneous, \cite{hirabayashi2015internal} calculated the analytical solutions of the stress field within the asteroid, and thus predicted the region most prone to failure based on the yield criterion.

Here we apply their methodology to estimate the failure mode of 2016 HO$_{3}$.
The complete elastic solution of a gravitational spinning ellipsoid in equilibrium has been obtained in \cite{dobrovolskis1982internal} by solving the elastic equations, including the equilibrium equations, the constitutive equations (Hooke's law), the strain–displacement equations, and zero-traction boundary conditions.
Figure \ref{fig:theoryEvolution} shows our stress analyses for the 2016 HO$_{3}$ Shape I model in various spin periods.
It is clear that the central region is under intensified pressure, and with the increase of spin velocity, there is a notable decline in pressure, ultimately leading the entire body into a tensile state.
The maximum tensile stress is still situated at the center.

Based on the elastic stress field above, we can then integrate the equilibrium stress solution with the failure criterion to predict the failure mode of a cohesive body.
\cite{zhang2018rotational,sanchez2012simulation,sanchez2018rotational} used the Drucker–Prager yield criterion (D-P criterion), a pressure-shear dependent yield criterion, which is given as
\begin{equation}
    \sqrt{J_2} \leq C + {2 \sin \theta} \/ [\sqrt{3} (3 - \sin \theta)] I_1
\end{equation}
Here $I_1$ is the first invariant of the Cauchy stress tensor, and $J_2$ is the second invariant of the deviatoric stress tensor.
$C$ and $\theta$ are the macro cohesive strength and internal friction angle of the body, respectively.
Therefore, for any point within the asteroid, its local strength must exceed a threshold $C_{\rm{need}} =  \sqrt{J_2} - {2 \sin \theta} \/ [\sqrt{3} (3 - \sin \theta)] I_1 $ to maintain shear stability.
Then $C_\mathrm{min}$, the minimal cohesion requisite for global structural integrity, can be determined by the maximum value of $C_{\rm{need}}$ across all points within the body.
Figure \ref{fig:theoryEvolution} shows that at extremely low rotational speeds, the most unstable region lies on the body's surface. 
However, as the rotation speed increases, the area prone to failure migrates to the center, and the requisite minimal cohesion increases accordingly.

However, when the cohesion is significantly large, granular materials undergo a transition from plastic to brittle behavior \citep{cheng2022numerical}, making the Maximum Tensile Stress criterion (MTS criterion) particularly relevant for small bodies.
This criterion posits that a material will fail by tensile failure when the maximum principal stress $\mid{\sigma}_{3}\mid$ surpasses the material's tensile strength $C$. Mathematically, this is represented as
\begin{equation}
    \mid{\sigma}_{3}\mid \leq C
\end{equation}
Therefore, for any point within the asteroid, its local strength must exceed a threshold $C_{\rm{need}} =  {\mid \sigma}_{3} \mid$ to maintain tensile stability.
Then $C_\mathrm{min}$, the minimal cohesion requisite for global structural integrity, can be determined by the maximum value of $C_{\rm{need}}$ across all points within the body.
Figure \ref{fig:theoryEvolution} shows that at extremely low rotational speeds, the most unstable region lies on the body's surface. 
As the rotation speed increases, the area prone to failure migrates to the poles, and then to the center. 
At the same time, the requisite minimal cohesion also increases.
Surprisingly, $C_\mathrm{min}$ derived by the MTS criterion consistently exceeds the value attained by the D-P criterion at high spin rates; conversely, the opposite is true at low spin rates. 
This means at high spin rates, the granular assembly is more prone to tensile failure rather than shear failure.
Therefore, previous works \citep[e.g.][]{zhang2018rotational,hirabayashi2015internal} that solely considered the D-P failure criterion might systematically underestimate the minimal cohesion strength required for self-rotating asteroids to maintain their structural stability.
This will be validated by direct granular simulations in the next section.

Additionally, while the continuum approach deviates from the discrete structure of rubble pile asteroids, the critical strength it provides can serve as a precise reference for the critical strength of 2016 HO$_3$ assuming a monolithic structure. This will be employed in the Discussion section to infer the potential structure of 2016 HO$_3$.

\subsection{Critical cohesion strength and failure behavior}
\label{sec:criticalcohesion}
While the continuum mechanics approach can provide an estimation of the failure mode of cohesive asteroids, it is hard to capture the discrete nature of realistic rubble-pile asteroids.
In this section, we perform a series of granular simulations for the Shape I rubble-pile model with various cohesion strengths to explore the failure conditions and behaviors of 2016 HO$_{3}$. 

We start the spin-up simulation with an initial spin rate of $1\times10^{-4}$ rad/s, and then continuously apply angular momentum increments at a low constant speed, making the body to spin up with a rate of $2\times10^{-9}$ $\mathrm{rad/s^2}$ as shown in Fig.~\ref{fig:spinratepath}.
The increasing centrifugal forces would destabilize the asteroid at a critical spin period.
Figure \ref{fig:evolution0.006Pa} presents the evolution of the shape, pressure, packing ratio and scaled non-affine strain during the spin-up process of the 0.006 Pa body.
At a spin period of 5 h, the central region is under intensified pressure, and its magnitude even shows quantitative consistency with that predicted by the continuum theory in Fig.~\ref{fig:theoryEvolution}.
As the spin rate increases, there is a notable decline in pressure throughout the body, but the central region is still in compression.
At this stage, the system experiences almost no deformation (or only minor elastic deformation), so its angular velocity closely follows the predetermined linear growth path.
When the spin period reaches 4.32 h, the internal regions of the asteroid violate the D-P criterion, leading to shear failure.
This is clearly observed by the slip surface in Fig.~\ref{fig:evolution0.006Pa}, where the porosity shows a significant decrease compared to the surrounding areas due to the so-called shear dilatancy of granular materials.
Outside this range, the plastic deformation indicator $D_\mathrm{min}/r$ is very small, approaching zero at both ends of the asteroid. 
This further supports the notion that the asteroid only undergoes shear failure at the slip surface, while the rest of the body retains its original structure.
The deformation leads to spin-down of the aggregate to maintain the conservation of angular momentum.
Because only local shear failure occurs in such situations, the system can regain stability and then enter into the subsequent acceleration-failure-deceleration cycle, as shown by the undulating pattern observed in the red curve in Fig.~\ref{fig:spinratepath}.
After the accumulation of several localized shear failures, the 0.006 Pa body finally undergoes a global collapse and disintegrates into a binary system upon reaching its spin limit.
Subsequently, its spin rate experiences a steep decline and fails to regain stability.

The 0 Pa body behaves almost the same as the 0.006 Pa body during its spin-up process, except that the former has a smaller critical spin rate compared to the latter.
However, when increasing the cohesive strength to 0.032 Pa, the evolution of the rubble pile progresses in different manners (Fig.~\ref{fig:evolution0.032Pa}).
Higher cohesive strengths make the rubble-pile structure capable of preserving its morphology and internal packing efficiency at elevated spin rates. 
Concurrently, the magnitude of deformation preceding the final catastrophic failure diminishes, followed by a more abrupt disintegration.
The distribution of the plastic deformation indicator $D_\mathrm{min}/r$ clearly shows that the asteroid fractures into two halves at the fracture surface, while the rest of the body retains its original structure.
In such case, its spin rate directly enters into a steep decline stage as shown in Fig.~\ref{fig:spinratepath}, without any deceleration-stabilization process (corresponding to localized deformation).
Increasing the cohesive strength to 3.189 Pa makes no change to the failure type (the body still fractures into two halves at its central region as shown in Fig.~\ref{fig:evolution3.189Pa}), except for a remarkable increase in the failure spin rate.
This suggests that the rubble-pile asteroid has undergone a different mode of failure at higher cohesive strengths as opposed to lower strength regimes, which is reminiscent of the ductile-to-brittle transition of granular regolith on Phobos as found by \cite{cheng2022numerical}.
Note that contrary to the findings in \cite{zhang2018rotational}, we do not observe any dependency between the number and size of fractured fragments and the cohesive strength. 
This discrepancy may be attributed to the highly elongated shape of the asteroids simulated in this study, as opposed to the near-spherical morphology employed in their work.

Figure \ref{fig:cmin_model1} summarizes the minimum stable cohesion strength $C_\mathrm{min}$ against spin period measured by DEM simulations and predicted by theoretical methods.
It is clear that $C_\mathrm{min}$ derived by the MTS criterion consistently exceeds the value attained by the D-P criterion at high spin rates; conversely, the opposite is true at low spin rates. 
This means at high spin rates, the granular assembly is more prone to tensile failure rather than shear failure.
The critical point between the two phases occurs at approximately 0.01 Pa.
Such a phase transition aligns well with the observations made in DEM simulations (Fig.\ref{fig:evolution0.006Pa}-\ref{fig:evolution3.189Pa}): the 0.006 Pa asteroid undergoes shear failure, accompanied by multiple fluctuations in its spin rate curve before the final decline; in contrast, the asteroids with cohesive strength of 0.032 Pa or larger experience only an abrupt tensile disruption.
Consequently, the MTS criterion and the tensile failure mode emerge as superior paradigms for investigating the failure behavior of high-strength asteroids like P/2013 R3 \citep{jewitt2014disintegrating}.
Previous works that exclusively employed the D-P failure criterion have invariably led to an underestimation of the minimal cohesive strength requisite for the structural stability of self-rotating celestial bodies.
Additionally, the results predict that the asteroid 2016 HO$_{3}$ with Shape I should have a minimum cohesion of 8.34 Pa considering its spin period of 28 min.

\subsection{Dependence on asteroid morphology}
Based on the limited light curve data of 2016 HO$_{3}$, three possible triaxial shape models were derived for this asteroid in \cite{li2021shape}. 
In this section, we employ these three shape models as shown in Fig.~\ref{fig:morphology} to analyze the influence of the length-to-width ratio on its structural stability.

\begin{figure}
	\includegraphics[clip,trim=0mm 0mm 0mm 0mm,width=\columnwidth]{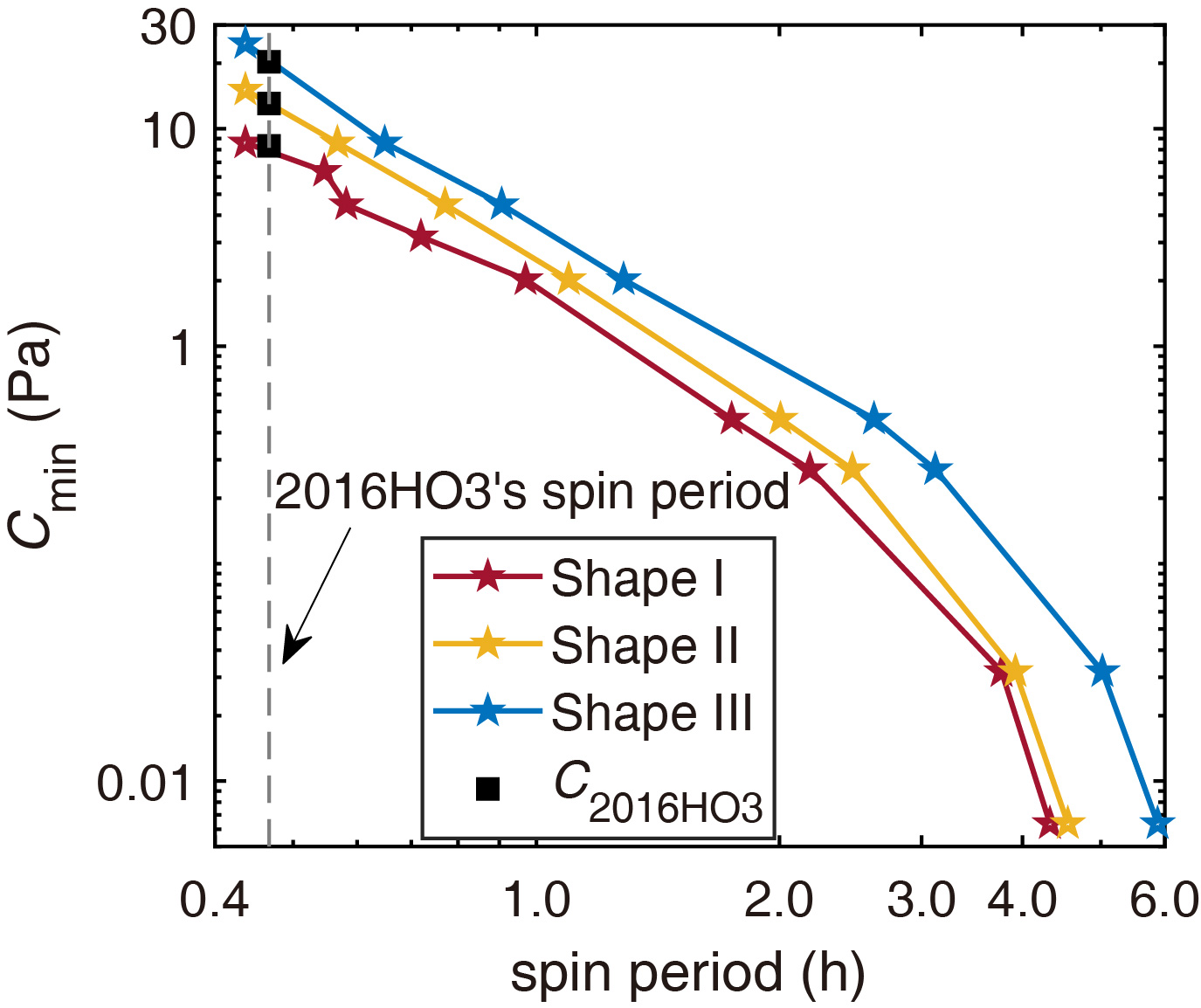}
    \caption{Minimum cohesive strength required for global structural integrity $C_\mathrm{min}$ against spin period for Shape I (red line), Shape II (yellow line) and Shape III (blue line) rubble-pile models.
    The black squares stand for the minimum cohesion for asteroid 2016 HO$_{3}$ with spin period of 28 min.}
    \label{fig:cmin_shapes}
\end{figure}

Our simulation results show that, regardless of how large the aspect ratio is ($c/a$ from 0.214 to 0.475 explored here), the failure behavior of the asteroid shows similar trends; that is, at higher cohesion strengths, the aggregate is more prone to tensile failure rather than shear failure upon spin-up.
And the minimum required cohesion $C_\mathrm{min}$ significantly increases with the spin rate (Fig.~\ref{fig:cmin_shapes}).
Furthermore, an increase in length-to-width ratio leads to a commensurate increase in the $C_\mathrm{min}$.
Specifically, for a body spinning with a period of 28 min, the minimum stable cohesion increases from 8.34 Pa for Shape I ($c/a = 0.475$), to 13.06 Pa for Shape II ($c/a = 0.333$), and to 20.31 Pa for Shape III ($c/a = 0.214$).
This suggests that elongated shapes are more susceptible and struggle to resist centrifugal forces.

\begin{figure*}
\centering
	\includegraphics[clip,trim=0mm 0mm 0mm 0mm,width=0.9\textwidth]{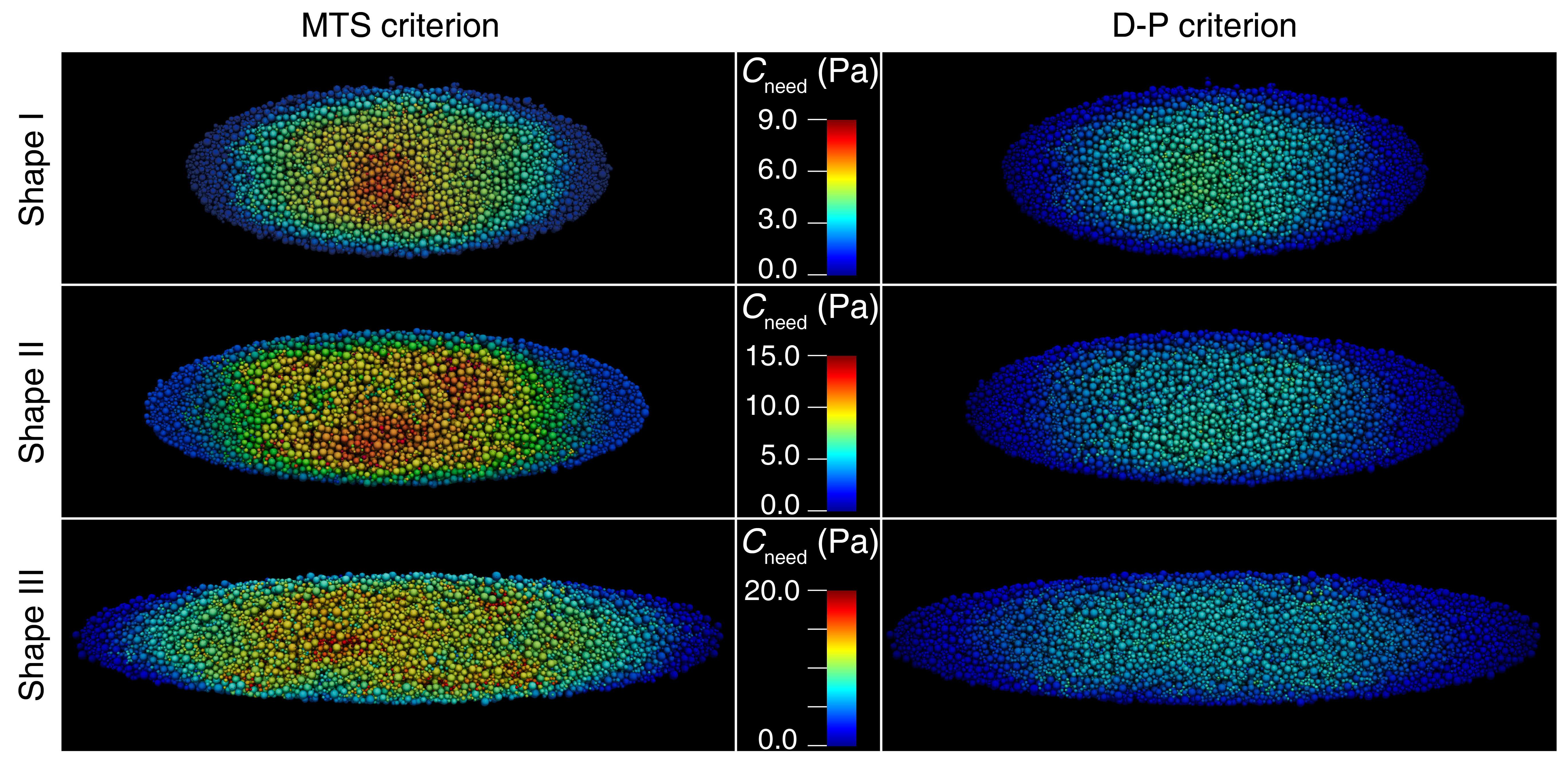}
    \caption{Distribution of the minimum required cohesion for keeping each local domain stable $C_\mathrm{need}$ over a cross section parallel to the spin axis for the three shape models at a spin rate of $4\times10^{-3}$ rad/s. 
    The left panel corresponds to the MTS criterion while the right one corresponds to the D-P criterion.
    A friction angle of $25^\circ$ is used in the derivation based on the D-P criterion.
    Note that $C_\mathrm{need}$ refers to the minimal cohesion strength required to sustain local stability, while $C_\mathrm{min}$ denotes the minimal cohesion strength essential for the global structural integrity of the asteroid, which equals to the global maximum of $C_\mathrm{need}$.}
    \label{fig:shapeanalysis}
\end{figure*}

\begin{figure*}
	\includegraphics[clip,trim=0mm 0mm 0mm 0mm,width=\textwidth]{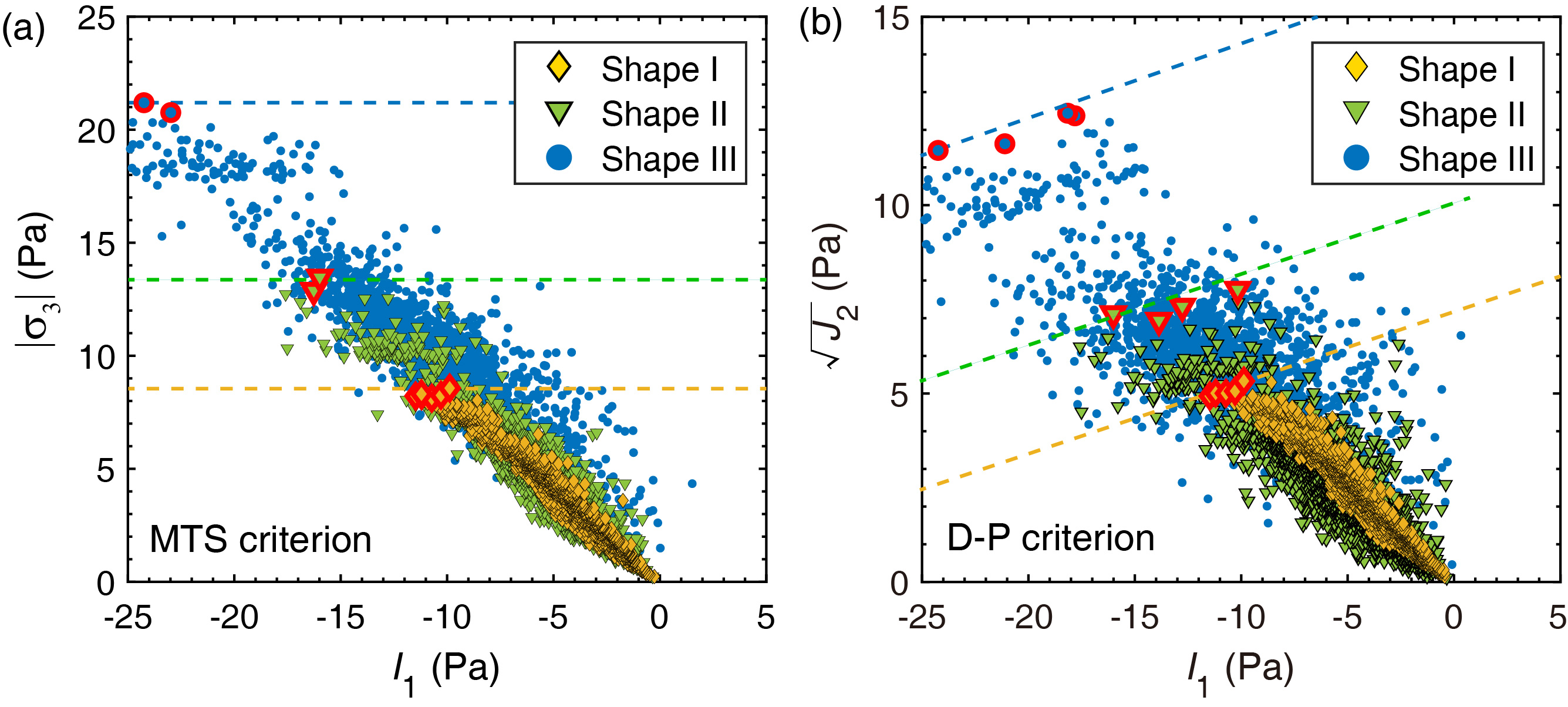}
    \caption{${\mid \sigma}_{3} \mid$ vs. $I_1$ (a) $\sqrt{J_2}$ vs. $I_1$ (b) for each particle within the rubble-pile models with three shapes at a spin rate of $4\times10^{-3}$ rad/s.
    Each symbol represent a particle whose $N_p$, the number of particles inside its coarse-grained domain, is larger than 200.
    This value is chosen to not only properly reveal the local stress distribution, but eliminate the large fluctuations of contact forces in the critical stage.
    The results of different shapes are shown in different colors as indicated in the legend.
    The failure points are highlighted by red symbols.
    The dashed lines stands for the failure envelopes as described by the MTS criterion (a) and the D-P criterion (b).
    }
    \label{fig:shapeanalysislaw}
\end{figure*}

To elucidate the mechanical origins of the observed shape dependency, we run a series of supplementary simulations and analyze the failure stress distribution within the body at a spin rate of $4\times10^{-3}$ rad/s.
Recall that through the continuum mechanics, for any point within the asteroid, its local strength must exceed a threshold $C_{\rm{need}} =  \sqrt{J_2} - {2 \sin \theta} \/ [\sqrt{3} (3 - \sin \theta)] I_1 $ to maintain shear stability (D-P criterion), and also must exceed a threshold $C_{\rm{need}} =  {\mid \sigma}_{3} \mid$ to maintain tensile stability (MTS criterion).
Note that $C_\mathrm{need}$ here refers to the minimal cohesion strength required to sustain local stability, while $C_\mathrm{min}$ denotes the minimal cohesion strength essential for the global structural integrity of the asteroid, which equals to the maximum of $C_\mathrm{need}$ across all points within the body.
Based on the spatial coarse-grained approach proposed in Section \ref{sec:micro2macro}, we calculate the macroscopic stresses and the corresponding $C_{\rm{need}}$ distributions across the asteroid body as shown in Fig.~\ref{fig:shapeanalysis}.
A friction angle of $25^\circ$ is used in derivation based on the D-P criterion.
The region most susceptible to failure, i.e., with largest $C_{\rm{need}}$, is located at the central area of the asteroid, matching up well with the location of the actual failure region found in DEM simulations.
Interestingly, $C_{\rm{need}}$ derived from the MTS criterion consistently exceeds that from the D-P criterion, which means the body is more prone to tensile failure rather than shear failure, again consistent with the actual failure mode.

Figure \ref{fig:shapeanalysislaw} shows the stress-state variable distribution for particles of the rubble-pile models with three shapes at their respective critical spin rate.
Panel (a) is ${\mid \sigma}_{3} \mid$ vs. $I_1$, and panel (b) is $\sqrt{J_2}$ vs. $I_1$.
Note that only particles with $N_\mathrm{neigh}$—the number of particles inside its coarse-grained domain—larger than 200 are considered.
This threshold is chosen to properly capture the local stress distribution while simultaneously eliminating the effect of large fluctuations within the force networks.
In MTS criterion, the failure points are identified as the ones that have the highest tensile stresses ${\mid \sigma}_{3} \mid$, i.e., the red symbols in Fig.~\ref{fig:shapeanalysislaw}(a). Then, the values of $C_{\rm{min}}$ can be determined from the best-fit failure envelopes, i.e., the dashed lines in the same figure.
Similarly, the failure points and the corresponding failure envelopes in D-P criterion are determined by the red symbols and dashed lines in  Fig.~\ref{fig:shapeanalysislaw}(b).
Clearly, $C_{\rm{min}}$ deduced from the D-P criterion are systematically lower than those deduced by the MTS criterion, which has already been validated in Fig.~\ref{fig:shapeanalysis}.
$C_{\rm{min}}$ deduced by the MTS criterion—which corresponds to tensile failure—even exhibit quantitative consistency with values found through direct simulations in Fig.~\ref{fig:cmin_shapes}.
Specifically, for the shape III model, the MTS criterion predicts $C_{\rm{min}}$ to be $\sim 21.25$ Pa, which is roughly aligning with $24.68$ Pa derived from direct simulations.
Therefore, the results here lend support for us to conclude that the tensile failure model and the MTS criterion are suitable to quantify the failure behavior of the rubble-pile asteroids with large cohesion strengths.
Additionally, we find the superfast rotator 2016 HO$_{3}$ should have a minimum cohesion of 8.34 Pa if considering shape I, 13.06 Pa if assuming shape II, and 20.31 Pa if considering shape III.
This provides important implications for addressing the structure of this body (see Section \ref{sec:discussion} below).


\subsection{Dependence on grain size ratio}
\begin{figure}
	\includegraphics[clip,trim=0mm 0mm 0mm 0mm,width=\columnwidth]{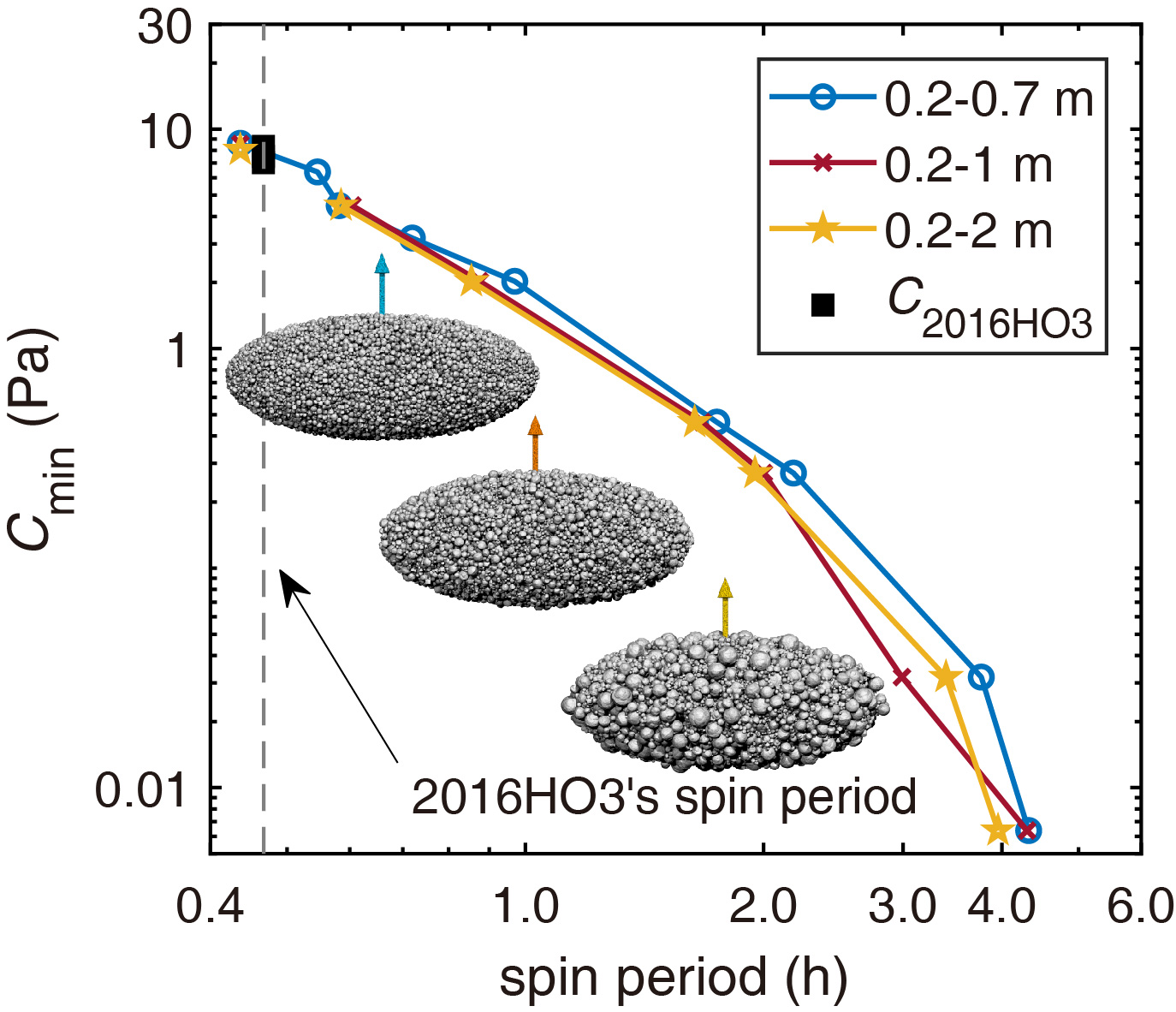}
    \caption{Minimum cohesive strength required for global structural integrity $C_\mathrm{min}$ against spin period for 0.2--0.7 m (blue line), 0.2--1.0 m (red line) and 0.2--2.0 (yellow line) rubble-pile models.
    The black squares stand for the minimum cohesion for asteroid 2016 HO$_{3}$ with spin period of 28 min.
    The insets show the rubble pile models with respective grain size ranges.}
    \label{fig:cmin_radius}
\end{figure}

We use the shape I model to explore the effect of constituent grain size.
As shown in Table \ref{tab:parameters}, we explore various grain size ranges including 0.2-0.7 m, 0.2-1.0 m and 0.2-2.0 m.
The ratio of the maximum grain size to the equivalent diameter ($\sim36$ m) of 2016 HO$_3$ spans from 0.020 to 0.028 to 0.056.
The last one is comparable to the boulders of approximately 25 m in size perched on the surface of asteroid Bennu (about 500 m in diameter).
For scenarios with lower cohesive strengths within the shear failure regime, the size distribution of constituent particles exerts a significant influence on the critical spin period as shown in Fig.~\ref{fig:cmin_radius}. Specially, for 0.032 Pa cases, the critical spin period increases from 3.0 to 4.0 h depending on the grain size ratio.
Conversely, when the cohesive strength of the asteroid is relative large to transition into the tensile failure regime, no discernible differences are observed for different grain size cases.
This is different from the finding by \cite{zhang2021creep} that the critical cohesion slightly increases with a lower particle resolution.
In their simulations, \cite{zhang2021creep} used a Didymos-like (near-spherical) aggregate with a spin period of 2.26 hr, and found the surface mass shedding serves as the predominant mode of failure.
In contrary, our simulated target is a highly elongate body in a super-fast rotating state ($\sim$28 min) and resides within the tensile failure regime.
The observed difference, therefore, may be due to the distinct features of the super-fast rotator 2016 HO$_3$.
This suggests that the morphology and rotational state of an asteroid could significantly influence its failure behavior. 
We should keep caution when extrapolating conclusions within a specific parameter range to other small bodies.
In total, we find the asteroid 2016 HO$_{3}$ should have a minimum cohesion of about 8.34 Pa regardless of its constituent grain size.


\subsection{Dependence on asteroid structure}
We use the shape I model to explore the effect of asteroid structure.
As shown in Table \ref{tab:parameters}, we investigate two rubble-pile configurations, the surface layer model and the boulder scatter model, in addition to the fine-grained model in the nominal cases.

\begin{figure*}
	\includegraphics[clip,trim=0mm 0mm 0mm 0mm,width=\textwidth]{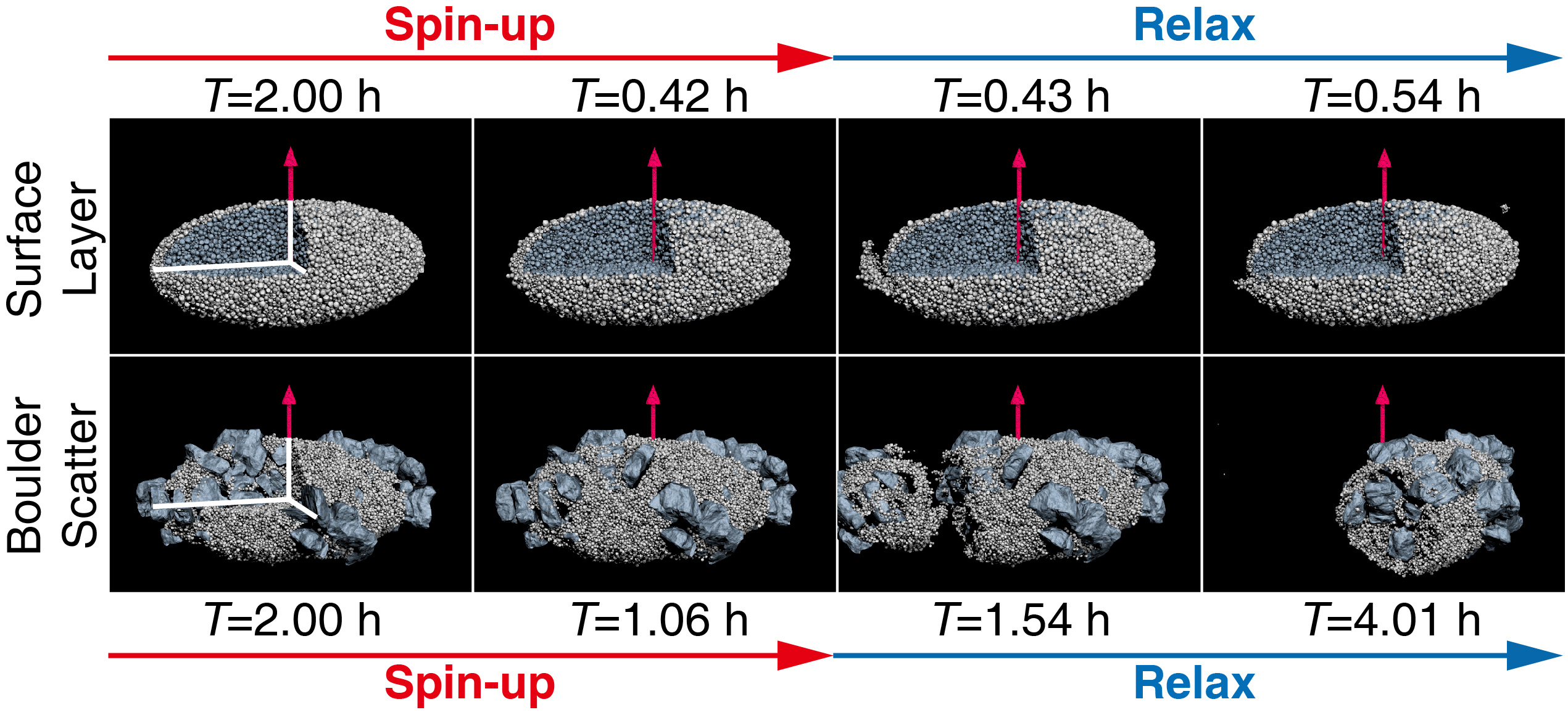}
    \caption{Evolution of the surface layer model asteroid and the boulder scatter model asteroid with the same regolith strength of 2.025 Pa. 
    The one with a surface loose layer is spun up to 0.42 h, at which time mass shedding failure occurs near the long axis of its equator. This leads to the body to slow down to 0.54 h due to the relaxation and generate a satellite. 
    The boulder-strewn one suffers tensile failure in its central area at a spin period of 1.06 h, and then slows down to 4.01 h due to the relaxation, forming an equal-mass binary system.}
    \label{fig:structureEvolution}
\end{figure*}

Figure \ref{fig:structureEvolution} illustrates the evolution of granular aggregates during the spin-up process, comparing the one with a loose surface layer to the other one interspersed with irregular boulders.
Both types possess loose regolith with a cohesive strength of 2.025 Pa.
Recall that the homogeneous body with the same strength in nominal cases typically fractures into two segments at its central region  when the spin period reaches 0.873 h.
For 2016 HO$_3$ with a layered structure, wherein loose regolith extends only to a depth of 2 m, its strong core acts to inhibit tensile failure in the central area.
Therefore, the failure zone is restricted solely to the superficial layer; that is, when the spin period of the body descends to 0.42 h, the outermost portion of the long axis of the surface regolith becomes dislodged from the main body, while the stronger interior remains substantially intact.
This mass shedding failure leads to a spin-down of the aggregate to maintain the conservation of angular momentum, and even generates a minimoon orbiting around the body, reminiscent of ubiquitous binary systems in asteroid belt \citep{richardson2006binary}.
In contrary, the evolution of the boulder scatter model progresses in different manners.
Despite having the same regolith strength as the surface layer model, its critical spin period significantly increases to 1.06 h, while strikingly close to the 0.970 h period of the fine-grained model.
Surprisingly, its failure mode also changes to a tensile failure near the central region, the same as the fine-grained model.
This implies the incorporation of boulders does not significantly influence the response of granular aggregates in critical states.
At such situations, the failure surface is a plane transverse to the long axis of the ellipsoid body.
Although the presence of boulders may disrupt the propagation of the failure surface, a complete failure surface can still be formed through interstitial low-strength regolith, thereby facilitating a tensile fracturing failure cleaving the rubble pile into two distinct segments.
Since the region of failure comprises the interstitial particles, it is the strength of these loose regolith, rather than that of the stronger boulders, which determines the threshold period for failure. 
Consequently, the failure period of the boulder scatter model closely approximates that of the fine-grained model, although some deviations may still arise.

Varying the cohesion strength of the loose regolith yields similar results.
In total, asteroids with the boulder scatter model have similar critical spin periods with the fine-grained model regardless of the cohesion strengths.
Additionally, the failure mode transitions from shear failure at low cohesive strengths to tensile failure at higher cohesive strengths, showing a similar trend with the fine-grained model.
However, there are still discrepancies between the two structures; the critical period of the boulder scatter model seems to systematically exceed that of the fine-grained model by less than 0.1 h, with the exception of the 0.462 Pa case where it is reversed.
We speculate that this may be due to the scattered distribution of irregular boulders within the rubble pile that disrupting the uniformity of the homogeneous structure.
For asteroids with a stiff interior covered by a loose regolith, the critical spin period is significantly reduced, i.e., the overall structure is more resistant to centrifugal forces.
Similarly, a failure mode transition is observed, where it changes from landslide deformation under lower strengths to tensile failure at higher strengths.
In total, we find the asteroid 2016 HO$_{3}$ should have a minimum cohesion of about 1.57 Pa if possessing a layered structure with a strong core, or about 8.18 Pa if possessing a boulder scatter structure.

\begin{figure}
	\includegraphics[clip,trim=0mm 0mm 0mm 0mm,width=\columnwidth]{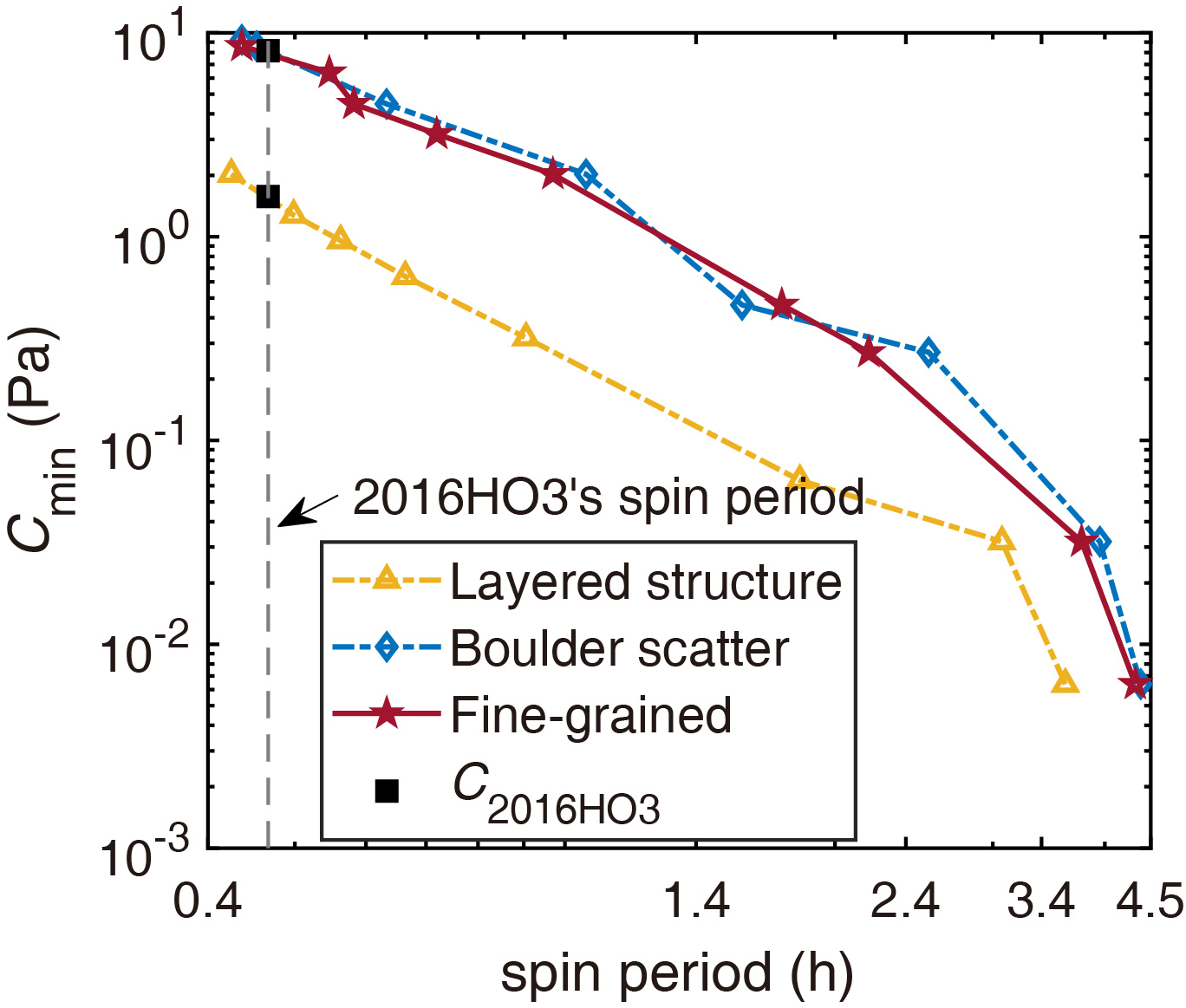}
    \caption{Minimum cohesive strength required for global structural integrity $C_\mathrm{min}$ against spin period for fine-grained structure (red line), surface regolith structure (yellow line) and boulder scatter structure (blue line).
    The black squares stand for the minimum cohesion for asteroid 2016 HO$_{3}$ with spin period of 28 min.}
    \label{fig:cmin_structure}
\end{figure}

\section{Discussion}
\label{sec:discussion}
Through the above parameter space study of the structural stability of asteroid 2016 HO$_3$, we determine that its superfast rotating state requires a minimum strength of about 7.0 Pa to 20.0 Pa to prevent tensile failure if possessing a fine-grained rubble-pile structure, regardless of its constituent grain size distribution.
The presence of giant boulders, even 10\% the size of the asteroid, within the aggregate would not significantly change the critical strength and failure mode.
However, if 2016 HO$_3$ has a stiff interior beneath a loose regolith layer, the minimum stable strength of the granular surface is significantly reduced to about 1.26 Pa.
Therefore, by comparing the minimum strength derived from this study for asteroid 2016 HO$_3$ with actual measured strength data from previous missions, we can infer the possible structure of this asteroid.

Spacecraft in situ observations and meteorite measurements have provided estimations of the potential range of asteroid strengths.
NEAR Shoemaker spacecraft observed widespread occurrence of tectonic landforms on S-type asteroid 433 Eros, which suggest its near-surface strength is from 1 to 6 MPa \citep{watters2011thrust}.
This supports the interpretation that the 16 km size asteroid Eros is a coherent body held together by material strength but with a pervasive fabric of fractures.
Another S-type asteroid, Itokawa, which has been closely explored by the Hayabusa spacecraft, is only about 300 m in size and was the first asteroid clearly identified as a rubble pile \citep{fujiwara2006rubble}.
Widespread imbricated boulders and particle sorting indicate that Itokawa has experienced global-scale landslide-like granular processes \citep{miyamoto2007regolith}.
Slope stability analysis shows that a cohesion of $\gtrsim 1.0$ Pa would significantly suppress surface mobility, which is inconsistent with the evidence for surface flow. 
Therefore Itokawa should has a week surface layer with a strength of mere $\lesssim 1.0$ Pa.
The Hayabusa2 and OSIRIS-REx missions have conducted detailed measurements of the strength of asteroids Ryugu and Bennu, respectively. Although both are C-type asteroids, their estimations still provide a reference for understanding the strength of the asteroid population including S-type asteroids.
Measurements of thermal properties have enabled estimation of the strength and porosity of discrete boulders on rubble-pile asteroids.
Thermal conductivity analysis by \cite{rozitis2020asteroid} distinguished the boulder population on Bennu into two types and gave tensile strengths of 0.10 to 0.15 MPa and 0.31 to 0.78 MPa for the low- and high-reflectance boulders, respectively.
This is comparable to the tensile strength (inferred from high porosities) of boulders on Ryugu, which ranges from 0.2 MPa to 0.28 MPa \citep{grott2019low}.
\cite{ballouz2020bennu} developed a novel framework to estimate the disruption threshold and impact strength of monolithic objects.
Through analyzing the maximum size and depth of craters observed on boulders on asteroid Bennu, they found the impact strength of metre-sized boulders is 0.44 to 1.7 MPa.
These values are similar to the compressive strength of 0.25–0.7 MPa estimated for the ungrouped C2 Tagish Lake fireball \citep{brown2002entry}.
Note that the above measurements are for discrete monolithic boulders.
The ensemble of particles (or regolith), at an asteroid’s surface or constituting the entire rubble pile asteroid, diminishes the significance of the strength of individual particles, instead makes the interactions among these discrete particles to determine the overall strength of the system.
On Ryugu, the Hayabusa2 mission’s artificial impact experiment indicated a weak surface with an effective strength of $\lesssim 1.3$ Pa \citep{arakawa2020artificial}.
The presence of an extensive ejecta deposit around the natural crater Bralgah on Bennu also implies a similar strength of surface material of $\lesssim 2$ Pa.
These values are orders of magnitude less than the cohesion of discrete boulders.
Nevertheless, they are supported by other lines of evidence.
\cite{lauretta2022spacecraft} found the size of the excavated region from the gas release of the OSIRIS-REx spacecraft's sampling operation requires a nearly cohesionless material ($\lesssim 1$ Pa).
The force at contact also suggested a strength of $\lesssim 1$ Pa in the near subsurface \citep{walsh2022near}.
Geotechnical stability analysis by \cite{barnouin2022formation} indicated a surface cohesion of $\lesssim 0.6$ Pa to explain the subtle latitudinal terraces on Bennu.
Not just the surface, but the entirety of Bennu should be with low strengths ($\lesssim 1.3$ Pa) to consistently account for all the geophysical features of Bennu and explain the absence of moons \citep{zhang2022inferring}. 

Clearly, the previous measurements of strength of asteroids fall into two distinct ranges: Monolithic bodies, be they surface discrete boulders or the coherent asteroids themselves, typically possess strengths in the megapascal range. In contrast, the strength of rubble pile structures is several orders of magnitude less, approximately within the pascal range.
Therefore, the minimum stable strength of asteroid 2016 HO$_3$ (tens of Pascals) is higher than previous estimates for the strength of rubble pile asteroids, which means its likelihood of being a typical rubble pile structure like Itokawa and Bennu is relatively low. 
However, we acknowledge that the disparity in strength is not markedly large, thus the potential for a rubble pile structure cannot be completely ruled out.
If asteroid 2016 HO$_3$ is indeed a small monolithic fragments generated by impacts on the Moon \citep{sharkey2021lunar} or on Main-belt asteroids, then the strength derived from prior empirical deductions would be completely sufficient to sustain its current fast rotation state. This implies that, from the perspective of structural stability, a monolithic structure is more plausible.
Despite this, over the span of several million years, thermal fatigue and meteoroid impacts could have altered the original monolithic structure, and produced fractures and grains on its surface.
The stability threshold for the loose surface layer ($\sim$1 Pa) falls within the range of measurements from previous missions, suggesting that asteroid 2016 HO$_3$ plausibly retains a layer of granular regolith, particularly in the stable regions near its poles.

In summary, the high rotational speed of asteroid 2016 HO$_3$ suggests a greater likelihood of a monolithic structure compared to a typical rubble pile structure, but the latter's possibility cannot be entirely dismissed. 
Nevertheless, the asteroid's surface could still retain a weak ($\sim$ Pa) regolith layer, especially near the polar regions.
This insight is valuable for designing sampling mechanism and selecting landing site for China's sample return missions to asteroid 2016 HO$_3$.


\begin{figure}
	\includegraphics[clip,trim=0mm 0mm 0mm 0mm,width=\columnwidth]{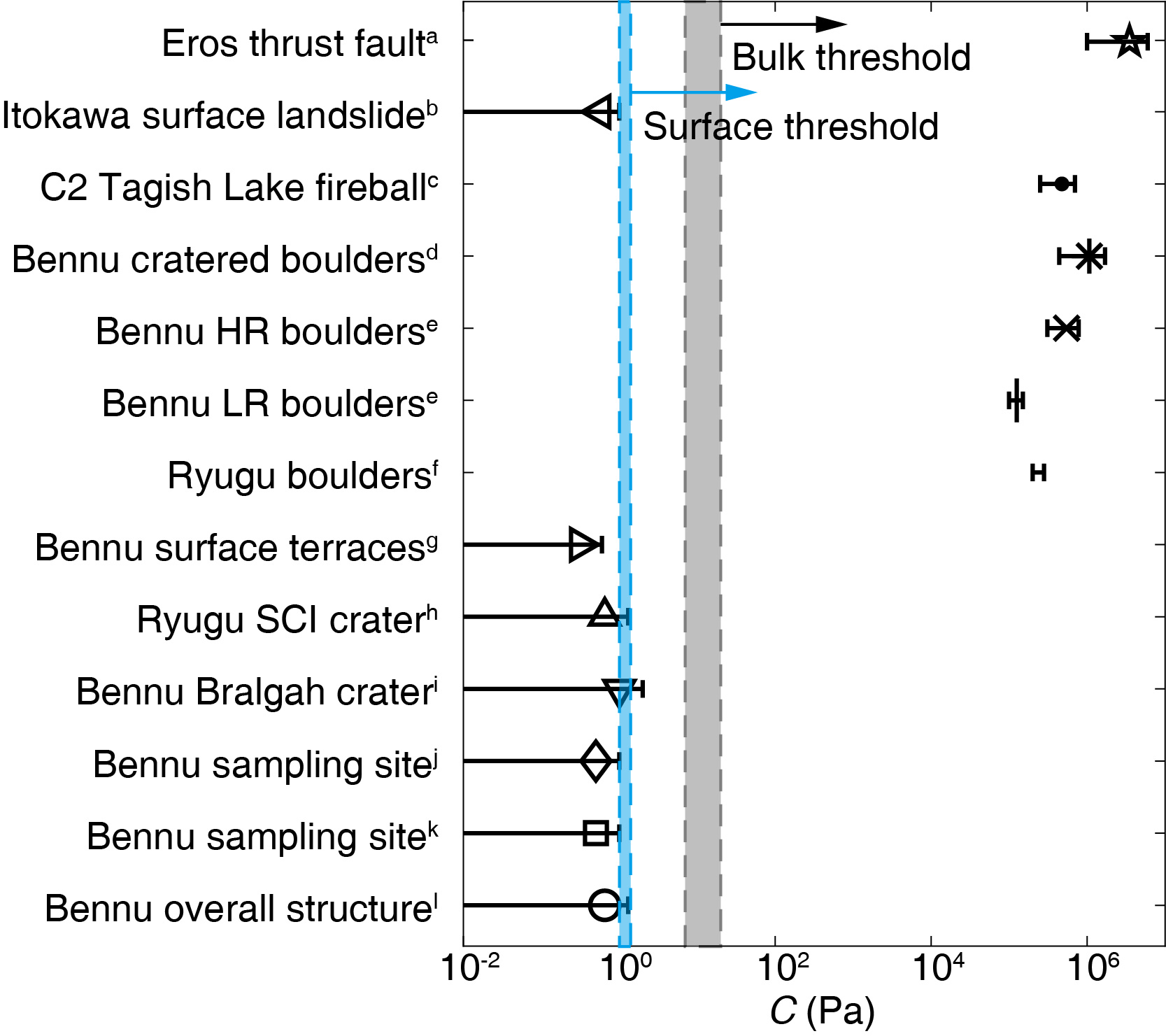}
    \caption{Estimations of asteroid strength and predictions for 2016 HO$_3$ strength. 
    (a) 1 to 6 MPa, estimated for the near-surface of Eros by \citet{watters2011thrust}. 
    (b) $\lesssim$ 1.0 Pa, for Itokawa’s granular surface as suggested by the regolith migration on its surface in \citet{miyamoto2007regolith}.
    (c) 0.25–0.7 MPa, estimated for the ungrouped C2 Tagish Lake fireball by \citet{brown2002entry}. 
    (d) 0.44–1.70 MPa, estimated for cratered boulders on Bennu by \citet{ballouz2020bennu}. 
    (e) 0.10 to 0.15 MPa and 0.31 to 0.78 MPa for the low- and high-reflectance boulders on Bennu, respectively, as reported by \citet{rozitis2020asteroid}. 
    (f) 0.2 MPa to 0.28 MPa, estimated for boulders on Ryugu by \citet{grott2019low}. 
    (g) $\lesssim$ 0.6 Pa, for Bennu’s granular surface as suggested by its terraces in \citet{barnouin2022formation}. 
    (h) $\lesssim$ 1.3 Pa, for Ryugu’s surface deduced from the SCI impact by \citet{arakawa2020artificial}. 
    (i) $\lesssim$ 2.0 Pa, for Bennu’s surface inferred from impact ejecta deposit according to \citet{perry2022low}. 
    (j) $\lesssim$ 1.0 Pa, estimated for the sampling site of Bennu by \citet{lauretta2022spacecraft}. 
    (k) $\lesssim$ 1.0 Pa, for Bennu's surface inferred from the touchdown force by \citet{walsh2022near}. 
    (l) $\lesssim$ 1.3 Pa, for the bulk strength of Bennu estimated from its structural history by \citet{zhang2022inferring}.
    The grey shaded region shows the minimum cohesive strength required for bulk structural integrity of 2016 HO$_3$, and the blue shaded region corresponds to the minimum cohesive strength required for its surface stability.
    }
    \label{fig:observation}
\end{figure}

\section{Conclusion}
\label{sec:conclusion}

Using the open-source soft-sphere discrete element code DEMBody, we study the rotational failure modes and behaviors of asteroid 2016 HO$_3$ with various morphologies, grain size distributions and structures. 
A bonded-aggregate approach is introduced that allows us to model a boulder-strew rubble pile, rather than being restricted to the sphere-based aggregates used in previous studies.

We find a 2016 HO$_3$ shaped rubble pile asteroid would undergone tensile failure at higher cohesive strengths as opposed to shear failure in lower strength regimes, regardless of its shape and constituent grain size ratio.
Such a failure mode transition is consistent with the priority between the MTS criterion and the D-P criterion as predicted by the continuum mechanics theory.
Therefore, previous works that solely considered the D-P failure criterion have systematically underestimated the minimal cohesion strength required for fast-rotating asteroids.
Increasing the aspect ratio of the asteroid's shape would significantly raises the critical strength, but altering the size distribution of the particles composing the rubble pile has minimal influence, even when inserting giant boulders that are approximately 10\% of the entire asteroid's size.

We predict that the high rotational speed of asteroid 2016 HO$_3$ requires a surface cohesion larger than $\sim$1 Pa and a bulk cohesion larger than $\sim$10--20 Pa.
Through comparing these strength conditions with the latest data from asteroid missions, we suggest a higher likelihood of a monolithic structure over a typical rubble pile structure. However, the possibility of the latter cannot be completely ruled out.
In addition, the asteroid's surface could still retain a loose regolith layer globally or only near its poles, which should be the target for sampling of China's asteroid mission.
Asteroids of tens of meters in size, yet to be explored by space missions, are among the least understood small celestial bodies. They are not only the most frequent potentially hazardous objects but could also be the most accessible space resources. 
China's Tianwen-2 mission is poised to test these structural hypotheses and will fill the research gap regarding small-sized asteroids within the NEA population.

\section*{Acknowledgements}

This work is supported by the National Natural Science Foundation of China under Grant 62227901.
B.C. is also supported by the National Natural Science Foundation of China (No. 12202227) and the Postdoctoral Innovative Talent Support Program of China (No. BX20220164). 

\section*{Data Availability}

The data underlying this article will be shared on reasonable request to the corresponding authors.



\bibliographystyle{mnras}
\bibliography{example} 



\bsp	
\label{lastpage}
\end{document}